\documentclass[amsmath,amssymb,amsbsy,reprint,prb,preprintnumbers,showpacs,superscriptaddress]{revtex4-2}
\usepackage{graphicx,color}
\usepackage{dcolumn}
\usepackage{bm}
\usepackage{braket}
\usepackage{mathtools}
\usepackage{ulem}
\usepackage[breaklinks,colorlinks=true,linkcolor=blue,urlcolor=blue,citecolor=blue]{hyperref}
\usepackage{times}
\usepackage{physics}
\usepackage{latexsym}
\usepackage{amsmath, amssymb}
\usepackage{mathtools}
\usepackage{multirow}

\newcommand{\be}{\begin{eqnarray}}
\newcommand{\ee}{\end{eqnarray}}

\newcommand{\nin}{\noindent}

\begin{document}

\title{Interplay between lattice gauge theory and subsystem codes}
\date{\today}
\author{Yoshihito Kuno} 
\affiliation{Graduate School of Engineering Science, Akita University, Akita 010-8502, Japan}
\author{Ikuo Ichinose} 
\thanks{A professor emeritus}
\affiliation{Department of Applied Physics, Nagoya Institute of Technology, Nagoya, 466-8555, Japan}


\begin{abstract}
It is now widely recognized that the toric code is a pure gauge-theory model governed by a projective Hamiltonian with topological orders.
In this work, we extend the interplay between quantum information system and gauge-theory model from the viewpoint of subsystem code, which is suitable for \textit{gauge systems including matter fields}.
As an example, we show that $Z_2$ lattice gauge-Higgs model in (2+1)-dimensions with specific open boundary conditions is nothing but a kind of subsystem code.
In the system, Gauss-law constraints are stabilizers, and
order parameters identifying Higgs and confinement phases exist and they are nothing but logical operators in subsystem codes residing on the boundaries.
Mixed anomaly of them dictates the existence of boundary zero modes, which is a direct consequence of symmetry-protected topological order in Higgs and confinement phases.
After identifying phase diagram, subsystem codes are embedded in the Higgs and confinement phases.
As our main findings, we give an explicit description of the code (encoded qubit) in the Higgs and confinement phases, which clarifies duality between Higgs and confinement phases.
The degenerate structure of subsystem code in the Higgs and confinement phases remains even in very high-energy levels, which is analogous to notion of strong-zero modes observed in some interesting condensed-matter systems. 
Numerical methods are used to corroborate analytically-obtained results and the obtained spectrum structure supports the analytical description of various subsystem codes in the gauge theory phases.
\end{abstract}


\maketitle
\section{Introduction}
Lattice gauge theory (LGT) \cite{Kogut1979} was invented to describes physical phenomena of elementary particle physics and succeeded in explaining quark-confinement phenomenon in strong interactions \cite{Wilson1974}. 
The application of the LGT is rich from condensed matter \cite{Fradkin_text,IchinoseMatsui2014} to quantum information \cite{Kitaev2003,Wang2003,Pachos2012}. 
In particular, some of quantum phases emergent in LGT are closely related with quantum memory \cite{Dennis2001,Kitaev2003,Arakawa2004,Ohno2004}. 
Their fault tolerance is explained by the notion of topological order \cite{Wen1990,QI_text} in condensed matter physics. 
Description of local-gauge symmetry in the LGT is also related to the notion of stabilizer in quantum information \cite{Nielsen_Chuang}. Gauss law of the LGT giving a strong constraint on the Hilbert space can be regarded as stabilizer condition in quantum information theory \cite{Pachos2012}. 
It is also known that some states constrained by Gauss law in LGT models acquire ability of quantum-error correction \cite{Dennis2001,Kitaev2003}. 
This is done by restoring a distorted state back to the original one by using the Gauss-law constraint.  
(In this sense, deep understanding of LGT has the potential to lead us into some interdisciplinary discoveries.)

Recently, the lattice gauge-Higgs model \cite{Fradkin1979} has been revisited in some studies \cite{Borla2021,Verresen2022} from a viewpoint of condensed matter physics. 
It was suggested that Higgs phase can be regarded as a symmetry protected topological phase (SPT phase) \cite{Verrsen2017,Tasaki2020}, and in (2+1) dimensions ((2+1)-D) with \textit{cylinder boundary conditions}, the Higgs and confinement regimes are distinguishable by observing operators on 
boundaries \cite{Verresen2022}.
This observation is connected to concept of one-form symmetry inherent in 
gauge theory~\cite{McGreevy2022}, which clarifies the origin of 't Hooft  loop as well as Wilson loop.
In the aspect of quantum information science, Ref.~\cite{Wildeboer2022} gave us a hint to notice 
that lattice gauge-Higgs models with open boundaries can be a subsystem code \cite{Poulin2005}, that is, degenerate eigenstates of the lattice gauge-Higgs model behave as encoded qubits. 
This suggested an important fact that not only in the ground state but also almost all higher-energy states in the model correspond to qubits in the subsystem code, and they can be stable in time evolution with long periods.

\begin{figure}[t]
\begin{center} 
\includegraphics[width=6.5cm]{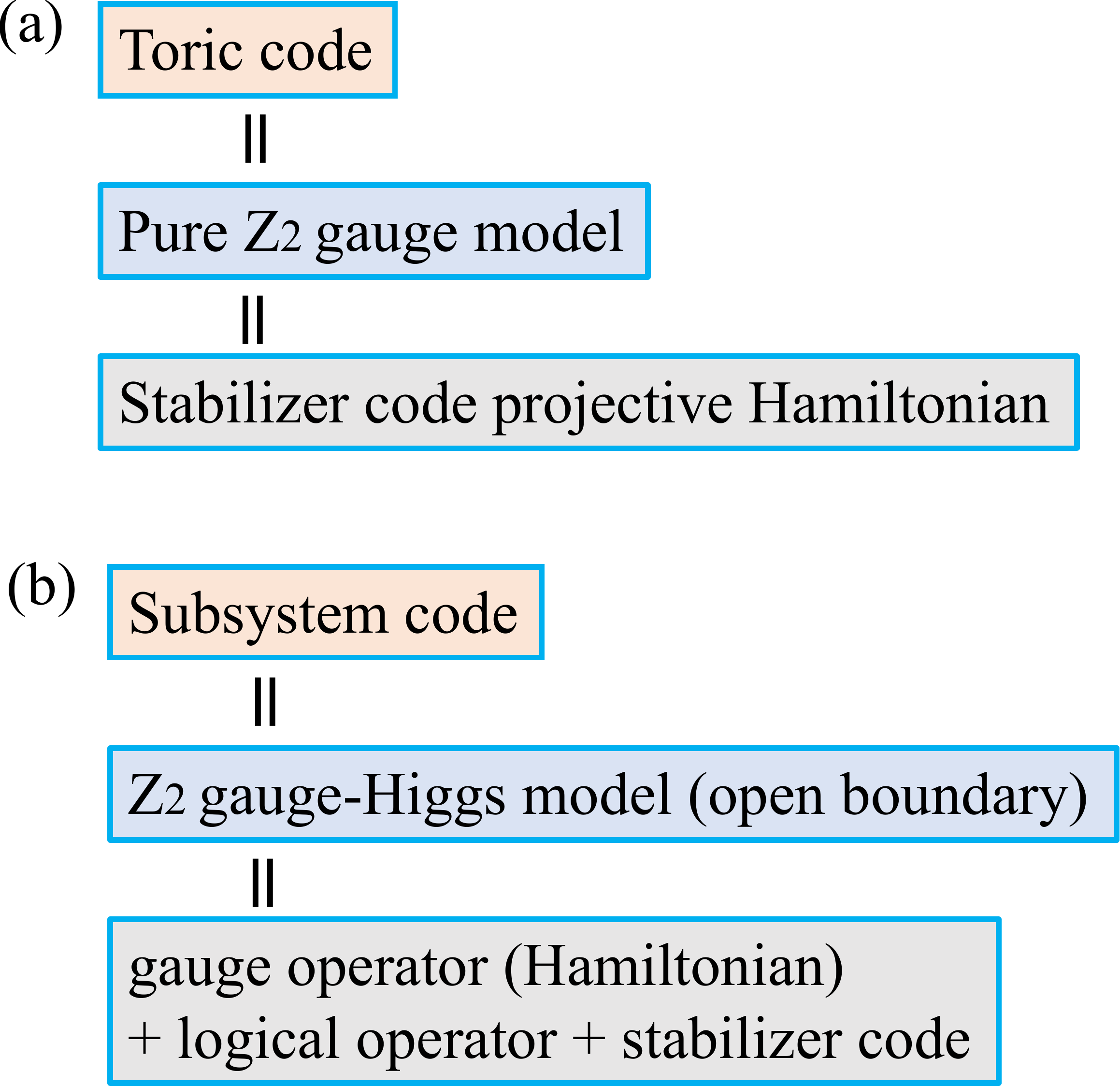} 
\end{center} 
\caption{Schematic of (gauge model)-(information code) relationship proposed in 
the present work: (a) Toric code and (b) Subsystem code.
The toric code is a projective pure gauge system, Hamiltonian of which is composed of stabilizers commutative with each other.
On the contrary, for ordinary gauge systems including matter degrees of freedom, all terms of Hamiltonian do not commute with each other.
For such a system, suitable quantum information notion is supplied by subsystem codes with gauge operators, logical operators, and stabilizers.
}
\label{Fig1}
\end{figure}

It is now widely recognized that the toric code is a pure gauge-theory model governed by 
a projective Hamiltonian with topological orders.
In this paper, we shall extend the interplay between quantum information system and gauge-theory model, in which the gauge degrees of freedom \textit{couple with matter fields}, and therefore, Hamiltonian 
\textit{is not projective}, i.e., all terms in the Hamiltonian do \textit{not} commute with each other.
There, the notion of subsystem code plays an important role.
(See Fig.~\ref{Fig1}.)
As a specific example of the above proposal, 
we take a further step towards detailed understanding of (2+1)-D lattice $Z_2$ gauge-Higgs models by following the previous works \cite{Verresen2022,Wildeboer2022}. 
To this end, we propose an extended version of the model with additional degrees of freedom of 
magnetic charge (flux), and also employ certain specific boundary conditions.
Although by a gauge fixing, the model reduces to the ordinary one, we can introduce suitable order 
parameters to clarify physical properties of the gauge-Higgs model. 

In the previous study \cite{Wildeboer2022}, a novel subsystem code was constructed.
This work raises interesting unresolved questions: 
(I) How the degenerate states of the subsystem code are understood in gauge-theory and SPT phase point of view?
(II) How `wavefunction' of the states for the subsystem code in each gauge-theory phase looks like? It is expected that the `wavefunction' clarifies physical meaning of SPT phase in gauge theory. 
(III) It is desired to obtain `wavefunctions' corresponding to multiple qubits, and high-energy states in the subsystem code, as they are expected to be closely related to strong zero modes discovered in Refs. \cite{Fendley2012,Fendley2016}.

In the following, we shall answer the above questions. 
To this end, we first clarify the relationship between the subsystem codes and the gauge-Higgs model. 
In particular, we explain that the logical operators of the subsystem codes are nothing but order 
parameters of the gauge-Higgs model, which are supplied by the employed open boundary conditions 
and play a central role in the present study. The spontaneous symmetry breaking (SSB) observed by the order parameters [SSB of charge and magnetic flux conservation symmetries]  clarifies phase diagram of the gauge-Higgs model. 
In this work, the explicit analytical descriptions of the degenerate encoded qubits in Higgs and confinement regimes are given. 
These descriptions elucidate an exact \textit{duality} between Higgs and confinement phases, and clarify physical properties of Higgs and confinement phases from the viewpoint of SSB. 
Duality in the present formalism reveals that the confinement phase is an SPT phase as the Higgs phase. Beyond single-encoded qubit, we shall give an explicit analytical description of general multiply-encoded qubits in Higgs and confinement regimes to show the utility of the correspondence between order parameters of the gauge theory and logical operator in quantum code. 
This is one of main findings in this work, corroborated by numerical studies.

We further focus on not only the ground-state multiplet but also on excited states in the system. As predicted by the recent work \cite{Wildeboer2022}, the degenerate spectrum structure of encoded qubit of subsystem code would be maintained in excited states. 
We shall verify this prediction exploiting the knowledge of the corresponding gauge theory. In particular, we find that the degenerate encoded qubits in Higgs and confinement regimes tend to survive in the entire spectrum, which is reminiscent of the strong zero mode discussed recently \cite{Fendley2012,Fendley2016}. 
To corroborate the analytical description for single and doubly-encoded qubits, we numerically investigate the degeneracy of encoded qubits as subsystem code in the entire energy spectrum of Higgs and confinement regimes. 
These results indicate that the subsystem codes with gauge-theory structure are generally robust up to high-energy regimes.

Before going into details of the model study, in order to capture the entire picture of the present proposal, we show schematically the relationship between the gauge-Higgs model and subsystem codes in Fig.~\ref{Fig1}.
We expect that this scheme is applicable for wide-range of lattice gauge models including matter fields, and is an extension of (toric code)-(gauge model) correspondence. The toric code is a pure $Z_2$ lattice-gauge model in $(2+1)$ D, and its Hamiltonian is composed of stabilizers, all of which are commutative with each other. This kind of system is sometimes called projective Hamiltonian system.
Figure~\ref{Fig1}(a) displays this fact.
Contrary to the toric code, the gauge-Higgs model is \textit{not} a projective system, i.e., all terms in 
the Hamiltonian do \textit{not} commute with each other. [See $H_{\rm GHM}$ in  Eq.~(\ref{HGHM}).]
For this case, a suitable notion in quantum information science is subsystem codes, which are composed of  
gauge operators, logical operators, and stabilizers \cite{Poulin2005}.
Please see Fig.~\ref{Fig1}(b). 
Hamiltonian of the gauge-Higgs model consists of the gauge operators instead of stabilizers,
whereas Gauss-law constraints play the role of stabilizers and the logical operators are order parameters.
Details of them will be explained in the main text.

The rest of this paper is organized as follows. 
In Sec.~II, we explain the target LGT model. 
We introduce Hamiltonian of (2+1)-D lattice $Z_2$ extended gauge-Higgs model and explain the specific boundary conditions employed in this work. 
We discuss symmetries and properties of the model, and then, derive an effective Hamiltonian by disentangling local-gauge degrees of freedom, which is an extension of the toric code with perturbations. 
Section III exhibits dramatic findings in this work.
Firstly, we explain notion of subsystem code (if the reader wants to know the essence of subsystem code in detail, see Refs.~\cite{Poulin2005,Wildeboer2022}), and then give detailed discussion on various properties concerning to gauge-theory ground state of the model.
In particular, physical motivation for introducing the open boundary and resultant SSB of the global charge symmetry are discussed. 
Secondly, we give the analytical description of the encoded qubit of subsystem code in Higgs and confinement regimes, and further discuss its extension to multiply-encoded qubit. 
Thirdly, in addition to the ground-state properties, we discuss degeneracy structure of excited states.
In Sec.~IV, we show the results of the numerical calculations, which corroborate analytical arguments given in Sec.~III. 
Detailed discussions on the numerical results and phase transition criticality are given. 
Section IV is devoted to discussion and conclusion.

Even though we use terminologies of quantum information, we hope that this article is readable for condensed matter, quantum information, and high-energy particle physicists.

\section{Model}
In this work, we shall study one of the LGT models, the gauge-Higgs model in two spatial dimensions.
Accessible review of the LGT is Ref.~\cite{Kogut1979}, and in particular for the gauge-Higgs mode, review is 
available in Ref.~\cite{IchinoseMatsui2014}.
As we explained in introduction, this gauge model is a good example for exemplifying  the subsystem-code formalism of the LGT.

\begin{figure}[t]
\begin{center} 
\includegraphics[width=8.5cm]{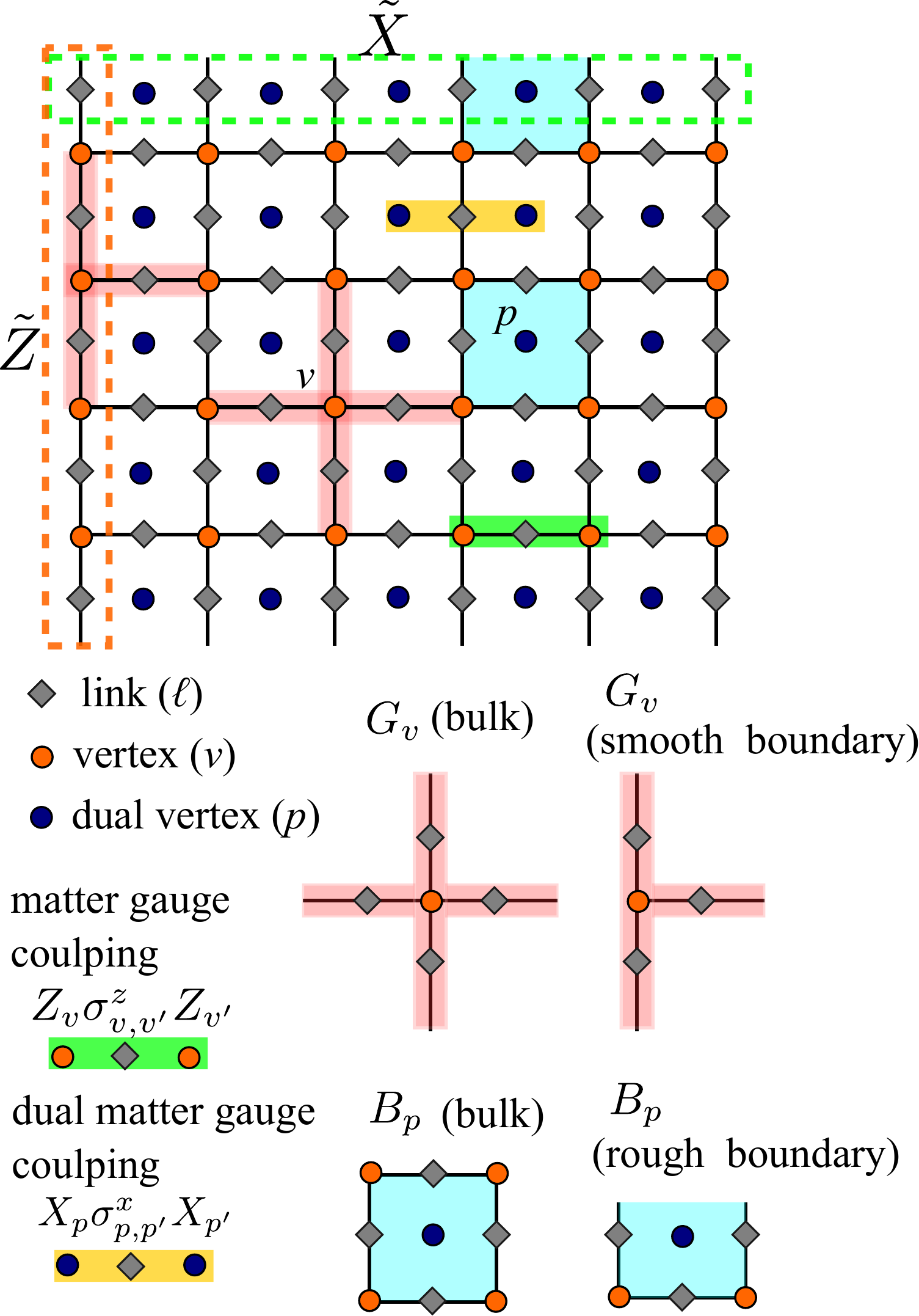}  
\end{center} 
\caption{Schematic figure of lattice with rough and smooth boundaries. 
The green dashed box represents the top rough boundary and the orange dashed box represents the left smooth boundary, on which logical operators are defined.}
\label{Fig2}
\end{figure}

We shall study a (2+1)-D $Z_2$ extended gauge-Higgs model, Hamiltonian of which is given by
\begin{eqnarray}
H_{\rm GHM}&=&\sum_{v}h_v X_v 
+\sum_{(p,p')}J^x_{p,p'}X_{p}\sigma^x_{p,p'}X_{p'}\nonumber\\
&+&\sum_{p}h_pZ_p+\sum_{(v,v')}J^z_{v,v'}Z_{v}\sigma^z_{v,v'}Z_{v'}.
\label{HGHM}
\end{eqnarray}
Here, we impose the following double gauge-invariant conditions, Gauss laws, for the physical subspace, 
\begin{eqnarray}
G_v|\psi\rangle=|\psi\rangle,\:\:
B_p|\psi\rangle=|\psi\rangle.
\label{const}
\end{eqnarray}
where
\begin{eqnarray}
&&G_v=X_v\prod_{\ell_{v} \in v} \sigma^x_{\ell_{v}}\equiv X_v\tilde{G}_v, \nonumber  \\
&&B_p=Z_p\prod_{\ell_{p} \in p} \sigma^z_{\ell_{p}}\equiv Z_p\tilde{B}_p,
\label{GvBp}
\end{eqnarray}
and $\ell_{v} \in v$ stands for links emanating from vertex (site) $v$, and 
$\ell_{p} \in p$ for links composing plaquette (box) $p$.
The $Z_2$-electric matter is defined on each vertex $v$, $(X_v,Z_v)$, and its magnetic dual,  $(X_p,Z_p)$, on each dual vertex $p$ 
(i.e., plaquette of the original lattice), where $X_v (Z_v)$ stands for 
the Pauli matrix $\sigma^x_v (\sigma^z_v)$, and similarly for  $X_p (Z_p)$ (See Fig.~\ref{Fig2}).
On the other hand, the $Z_2$ gauge field is defined on links and denoted by $(\sigma^x_\ell, \sigma^z_\ell)$, 
$\sigma^z_{v,v'}$ denote a gauge variable on link connecting  neighboring vertices $v$ and $v'$, 
and $\sigma^x_{p,p'}$ denote a gauge variable on link connecting neighboring dual vertices $p$ and $p'$. 
The gauge field $\sigma^x_\ell$ is related to the electric field $\hat{E}_\ell$ as $\sigma^x_\ell=e^{i\pi \hat{E}_\ell}$, and $\sigma^z_\ell$ to the vector potential $\hat{A}_\ell$ as $\sigma^z_\ell=e^{i\pi \hat{A}_\ell}$, and eigenvalues are $\{ 0, 1 \}$ for both the operators.
The electric field and vector potential are conjugate with each other, and operation of $\sigma^z_\ell$
produces electric flux on link $\ell$.

There are two differences between the system given by $H_{\rm GHM}$ [Eq.~(\ref{HGHM})] and  the ordinary 
$Z_2$ gauge-Higgs LGT: 
\begin{enumerate}
\item[(I)]
Dual matter field couples with the  gauge field and the  coupling term, $X_{p}\sigma^x_{p,p'}X_{p'}$ 
is added to the Hamiltonian besides the ordinary $Z_2$ electric matter-gauge coupling.
This $Z_2$ degrees of freedom residing on each plaquette $p$, $(X_p,Z_p)$, corresponds to `particle' carrying 
magnetic flux (magnetic charge), and its hopping induces fluctuation of the gauge field and, therefore, confinement of electric charges~\cite{Arakawa2004}. 
The ordinary electric term as well as the magnetic-plaquette term in the Hamiltonian are replaced with
dynamical variables $(X_p,Z_p)$, even though similar dynamics to the ordinary one emerges 
from $(X_p,Z_p)$.
\item[(II)]
By the presence of the magnetic-charge degrees of freedom, an additional local gauge symmetry emerges
such as, 
\begin{equation}
X_p \rightarrow X_p V_p, \; \; 
\sigma^x_{p,p'} \rightarrow  V_p\sigma^x_{p,p'}V_{p'},  \;\; V_p, V_{p'} \in Z_2,
\label{second}
\end{equation}
and we impose additional Gauss-law constraint on the physical state, Eqs.~(\ref{const}) and (\ref{GvBp}).
One may wonder that the system (\ref{HGHM}) reduces to the ordinary gauge-Higgs model by 
`integrating out' the magnetic charge degrees of freedom via, e.g., employing unitary gauge
of the second local gauge symmetry in Eq.~(\ref{second}).
In fact as we show shortly, disentangling of electric and magnetic particles generates the ordinary 
Hamiltonian of the gauge-Higgs model in unitary gauge.
However with specific open boundary conditions, which we shall employ, there emerges a small but 
important difference between the gauge-Higgs model (\ref{HGHM}) and the ordinary one.
\end{enumerate}

The reason why we employ the above Hamiltonian (\ref{HGHM}), which exhibits the electric-magnetic duality manifestly, as a starting model will become clear later on. 
The model of Eq.~(\ref{HGHM}) is different from the conventional toric code \cite{Kitaev2003}, 
in such a way that, the model of Eq.~(\ref{HGHM}) is not composed of stabilizers, and therefore, is not solvable, while the toric code is 
projective and solvable since all terms in the Hamiltonian are commutative with each other, i.e., stabilizers (See Fig.~\ref{Fig1}).

Here, we also note that the above Hamiltonian was recently proposed for describing subsystem quantum code producing fault-tolerant qubit \cite{Wildeboer2022}, 
although the explicit form of $H_{\rm GHM}$
in Eq.~(\ref{HGHM}) was not shown. 
There, each term of the Hamiltonian (\ref{HGHM}) is categorized as ``gauge'' operator in the subsystem-code literature \cite{Poulin2005}. 
[It is often remarked that the terminology ``gauge'' is rather confusing as it 
has nothing to do with gauge symmetry in LGT. See later discussion.]

We introduce a square lattice shown in Fig.~\ref{Fig2}, with specific boundaries named rough and smooth boundaries. 
For both boundaries, the form of $G_v$ and $B_p$ are explicitly shown, both of which are composed of three links with a vertex and a plaquette, respectively. 
Some of the electric-matter-gauge and magnetic-matter-gauge terms in the Hamiltonian, $H_{\rm GHM}$ in Eq.~(\ref{HGHM}), are missing on the boundaries. 
That is, the electric (magnetic) hopping on the rough (smooth) boundary does not exist.
We note that the exact electric-magnetic duality holds in the system. 

The model with the above open boundary conditions has four important symmetries, generators of which are given by,
\begin{eqnarray}
P&=&\prod_{v}X_v,\:\:
S_Z=\prod_{p}Z_p,\\
W_{\gamma}&=&\prod_{\ell \in \gamma}\sigma^z_{\ell},\:\:
H_{\gamma}=\prod_{\ell \in \gamma}\sigma^x_{\ell}.
\label{symmetry}
\end{eqnarray}
$P$ is the parity of the total electric charge, corresponding to the global spin flip on each vertex,
and similarly, $S_Z$ is the parity of the total magnetic flux per plaquette.
The boundary hopping terms are forbidden by the $P$ and $S_Z$ symmetries. 
$P$ and $S_Z$ are global topological symmetries, whereas $W_{\gamma}$ and $H_{\gamma}$ are one-form symmetry, which has been extensively studied recently \cite{Gaiotto2015, McGreevy2022}.
Although the $J^x_{p,p'}$ and $J^z_{v,v'}$ terms in the Hamiltonian $H_{\rm GHM}$ [Eq.~(\ref{HGHM})] explicitly break 
the one-form symmetries, it was shown that the higher-form symmetry is generally robust and 
give non-trivial effect on dynamics of the system. 
In the $(2+1)$-D system, $H_\gamma$ can be regarded as  't Hooft loop (string) dual to Wilson loop (string) $W_\gamma$.
In particular, the path $\gamma$ in the one-form symmetry $W_{\gamma}$ ($H_{\gamma}$) is arbitrary. 
It is easily verified that by using the Gauss laws in Eq.~(\ref{GvBp}),  $P$  ($S_Z$) is expressed as a 't Hooft 
(Wilson) `loop' residing on the top and bottom rough (left and right smooth) boundaries (see Fig.~\ref{Fig2}),
which we call boundary-one-form operators hereafter.
This fact plays an important role in later discussion on phase diagram of the gauge-Higgs model.

In the context of the present model $H_{\rm GHM}$, the decomposed Hilbert space of $H_{\rm GHM}$ can be regarded as subsystem code \cite{Poulin2005}, 
and its basic discussion is reviewed in \cite{Wildeboer2022}. 
The Hilbert space of the system $H_{\rm GHM}$ are characterized by gauge, stabilizer, and bare logical operators. 
The Hilbert space is operated by logical encoded qubit and gauge qubit \cite{Wildeboer2022}. 
The structure of subsystem code may give an efficient error correction route as discussed by proposing simple examples \cite{Bacon2006}.
The stabilizer of the subsystem code 
is given by the projectors $\{G_v,B_p,P,S_z\}$,  ``gauge'' operators are those commute with the projectors, corresponding to each terms of $H_{\rm GHM}$, and logical operators $\{\tilde{X}, \tilde{Z}\}$ are shown later in Sec.III. 
It should be noted that all eigenstates of $H_{\rm GHM}$ is labeled by $G_v=B_p=+1$ due to the condition of Eq.~(\ref{const}) while $P,S_z=\pm 1$ for eigenstates. 
This situation is different from some conventional stabilizer model such as toric code \cite{Kitaev2003,Pachos2012} and cluster model \cite{Son2011,Son2012,Zeng2016}, etc., 
where eigenvalue of every stabilizers takes $\pm 1$ depending on eigenstates.

Here, to capture the physical properties of $H_{\rm GHM}$ more clearly, we consider the following unitary transformations (sometimes called circuit unitary transformation):
\begin{eqnarray}
U_v&=&H\biggl(\prod_{v}\prod_{\ell\in v} (CZ)_{v,\ell}\biggr)H,\\
U_p&=&H\biggl(\prod_{p}\prod_{\ell\in p} (CZ)_{p,\ell}\biggr)H,
\end{eqnarray}
where $H$ is the Hadamard transformation on each link and 
$(CZ)_{i,j}$ is a controlled $Z$-gate for the site $i$ and link $j$.
Applying the above transformation to $H_{\rm GHM}$, we obtain the following effective disentangled model
\begin{eqnarray}
&&U_vU_pH_{\rm GHM}(U_vU_p)^\dagger\equiv H_{\rm TC},\nonumber
\end{eqnarray}
\begin{eqnarray}
H_{\rm TC}&=&\sum_{v}h_v\tilde{G}_v  +\sum_{\ell \notin \mbox{rough}}J^z_{\ell}\sigma^z_{\ell}\nonumber\\
&+&\sum_{p}h_p\tilde{B}_p+\sum_{\ell \notin \mbox{smooth}}J^x_{\ell} \sigma^x_{\ell}.
\label{HTC}
\end{eqnarray}
The above circuit disentanglement corresponds to gauge fixing with unitary gauge, in which degrees of freedom corresponding to local gauge transformation are eliminated~\cite{noteGF}. 
We remark that the specific form of the $J^z$ and $J^x$-terms originates from the starting Hamiltonian 
$H_{\rm GHM}$ in Eq.~(\ref{HGHM}). 

The model $H_{\rm TC}$ is nothing but an extended system of toric code \cite{Kitaev2003} including local 
perturbations ($J^z$ and $J^x$-terms) and with rough and smooth boundaries.
$H_{\rm TC}$ with periodic boundary conditions and also $H_{\rm TC}$ having the full $J^z$ and $J^x$-terms
under the standard open boundary conditions have been studied in previous works from the viewpoint of quantum information.

\section{Gauge-theory phase diagram and subsystem quantum code}
The model of $H_{\rm GHM}$ and $H_{\rm TC}$ has three distinguishable gauge-theory phases, 
i.e., Higgs, confinement and deconfinement (toric code/topological) phases under the specific type of present open boundary conditions. 
In particular, recent work exhibited that the Higgs phase is an SPT phase protected by the symmetries $P$ and $W_{\gamma}$, 
and can be distinguished from the confinement phase by employing open boundaries \cite{Verresen2022}, contrary to the previous common belief \cite{Fradkin1979}. 
Furthermore, even in the presence of explicit breaking of the one-form symmetry $W_{\gamma}$, the SPT phase survives as long as the gap is open. 
This is an important contribution for understanding the gauge model complementing the seminar and influential study of gauge-Higgs model \cite{Fradkin1979}.   

In this work, we go a step further and obtain a concrete form of boundary states in Higgs and confinement 
phases, which play an essential role in understanding the gauge-theory phase diagram and properties of
the subsystem code proposed in \cite{Wildeboer2022}. 

Degeneracy of quantum states in the system with suitable open boundary conditions for subsystem code is governed by the encoded qubit  \cite{Poulin2005}.
Logical operators for the code embedded in the Hamiltonian $H_{\rm TC}$ are given by  (see Fig.~\ref{Fig2}),
\begin{eqnarray}
\tilde{X}=\prod_{\ell\in \mbox{top rough}} \sigma^x_{\ell},\:\:\:
\tilde{Z}=\prod_{\ell\in \mbox{left smooth}} \sigma^z_{\ell}.
\label{logical_OP_1Q}
\end{eqnarray}
It is easily verified that they anticommute with each other $\{ \tilde{X},\tilde{Z} \} = 0$ since single link is 
shared by each rough and smooth boundaries, and also both of them
commute with the Hamiltonians, 
$[H_{\rm TC}, \tilde{X}]=[H_{\rm TC}, \tilde{Z}]=[H_{\rm GHM}, \tilde{X}]=[H_{\rm GHM}, \tilde{Z}]=0$. 
Note that the operators $\tilde{Z}$ and $\tilde{X}$ are \textit{invariant} under the disentangling transformation $U_vU_p$. 
Thus, these operators act as logical operators in the both models $H_{\rm GHM}$ and $H_{\rm TC}$, as pointed out in \cite{Wildeboer2022}.

We can introduce counterpart of $\tilde{X}$ and $\tilde{Z}$ residing on the bottom-rough and
right-smooth boundaries, respectively, defined such as,
\begin{eqnarray}
\tilde{X}^\ast=\prod_{\ell\in \mbox{bottom rough}} \sigma^x_{\ell},\:\:\:
\tilde{Z}^\ast=\prod_{\ell\in \mbox{right smooth}} \sigma^z_{\ell}.
\end{eqnarray}
As we explained in the above, 
Gauss laws in Eq.~(\ref{GvBp}) relate the global symmetry operators $P$ and $S_Z$ with the above boundary operators as follows,
\begin{eqnarray}
    S_X\equiv P=\tilde{X}\tilde{X}^\ast, \;\; S_Z=\tilde{Z}\tilde{Z}^\ast,
    \label{SXSZ}
\end{eqnarray}
where we have introduced the notation $S_X$, and the above relation eloquently tells us emergence of the symmetry fractionalization \cite{Verrsen2017}.

As $S_X$ and $S_Z$ commute with each other and also with the Hamiltonian, the full space of state can be divided into subspaces with eigenvalue of $(S_X,S_Z)=(\pm 1,\pm 1)$. 
However, $S_X$ and/or $S_Z$ can be SSB, and in that case, $\tilde{X} (\tilde{Z})$ and $\tilde{X}^\ast (\tilde{Z}^\ast)$ are independent operators.
On the other hand as $\tilde{X} (\tilde{X}^\ast)$ and  $\tilde{Z} (\tilde{Z}^\ast)$ anticommute with each other, gapless modes emerge in the vicinity of boundary. 
It is also possible that one of the symmetries generated by $\tilde{X}$ or $\tilde{Z}$ is spontaneously
broken by the condensation of the other as they also play a role of order parameter.
Pattern of the SSB clarifies physical picture of gauge-theory phase, which has been masked by the local-gauge symmetry so far.

In the following, we discuss the gauge-theory phases by showing explicit form of boundary states for each phase of the gauge-Higgs model, which has not been given so far, and then we shall verify the analytical observation by using numerical methods, exact diagonalization (ED), where we employ useful numerical package \cite{Quspin1,Quspin2}.

\subsection{Gauge-theory phases: Spontaneous symmetry breaking and confinement}

In this subsection, we shall take a look at phase diagram of the present gauge model, $H_{\rm TC}$ and $H_{\rm GHM}$.
As we mentioned in the above, the phase diagram of $H_{\rm TC}$ with periodic boundary conditions was investigated and it was clarified 
that there are three `phases', deconfined-topological, Higgs, and confinement phases. 
The seminal work in Ref.~\cite{Fradkin1979} showed that the Higgs and confinement phases are connected without thermodynamic singularities, that is, 
they are adiabatically connected.
However, in the model in a semi-infinite cylinder geometry, the Higgs and confinement phases are
distinct in the behavior of the boundary excitation \cite{Verresen2022}. 
In what follows, we mainly focus on physical properties in the vicinity of the boundaries, as they are essentially related to the SPT and also subsystem code. 
In addition, we shall show that the investigation of states near boundaries clarifies the SSB of the charge symmetry, mechanism of quark confinement, and duality between them.
This is a reason why we employ specific boundary conditions.

In what follows, we mostly focus on the model $H_{\rm TC}$, and carry out numerical calculation for $H_{\rm TC}$ in Sec. IV since the model is simple and, 
nevertheless, keeps physical essence of $H_{\rm GHM}$.

We first consider the Higgs regime for large negative $J^z_\ell$ such as $-J^z_\ell \gg |J^x_\ell|, |h_v|, |h_p|$.
In this regime, link variables in the bulk are ordered as $\langle \sigma^z_\ell \rangle \sim 1$, and this phase is regarded sometimes trivial.
However, this long-range order (LRO) in the bulk indicates $\langle \tilde{Z}\rangle \neq 0$, 
as the three-link plaquette terms on the rough boundary shown in Fig.~\ref{Fig2} generate ferromagnetic interactions between $\sigma^z$'s on dangling links. 
The fractionalized-symmetry operator $\tilde{X}$ operates on edge states nontrivially in that state. 
In fact, the existence of the LRO in the 1D rough boundary is numerically verified as we show later on (See Fig.~\ref{Fig7}).
Therefore, states on the rough boundaries are approximately given as $|\uparrow \uparrow \cdots\uparrow\rangle$ or $|\downarrow \downarrow \cdots\downarrow\rangle$.
The operator $\tilde{X}(\tilde{X}^\ast)$ interchanges these states,
$|\uparrow \uparrow \cdots\uparrow\rangle \leftrightarrow|\downarrow \downarrow \cdots\downarrow\rangle$.
On the top and bottom boundaries, the above two states can be taken independently, and therefore, the ground state is expected to be four-fold degenerate.
Here, we should remark that the gauge operators $\sigma^z_\ell$ creates (flips in $Z_2$ case) electric flux on link $\ell$, and therefore, 
its non-vanishing expectation value on the boundary means strong fluctuations of electric field and the SSB of the charge symmetry.

\begin{figure}[t]
\begin{center} 
\includegraphics[width=8cm]{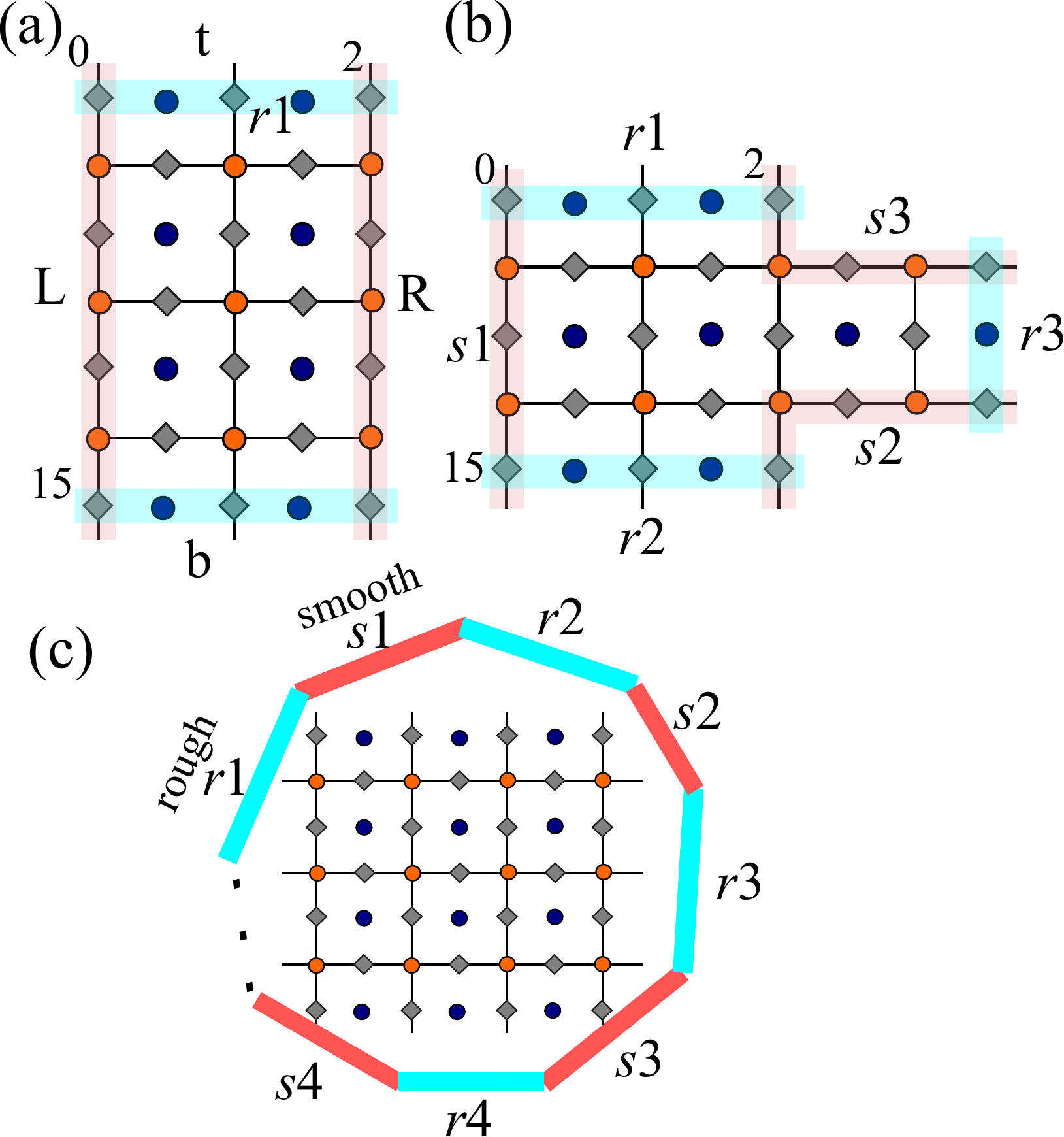}  
\end{center} 
\caption{(a) Single-encoded qubit system. The system has two rough and smooth boundaries.
(b) Doubly-encoded qubits system. The system has three rough and smooth boundaries. 
(c) $N$-encoded qubits system. The system has $N+1$ rough and smooth boundaries.
}
\label{Fig3}
\end{figure}

\begin{figure}[t]
\begin{center} 
\includegraphics[width=8cm]{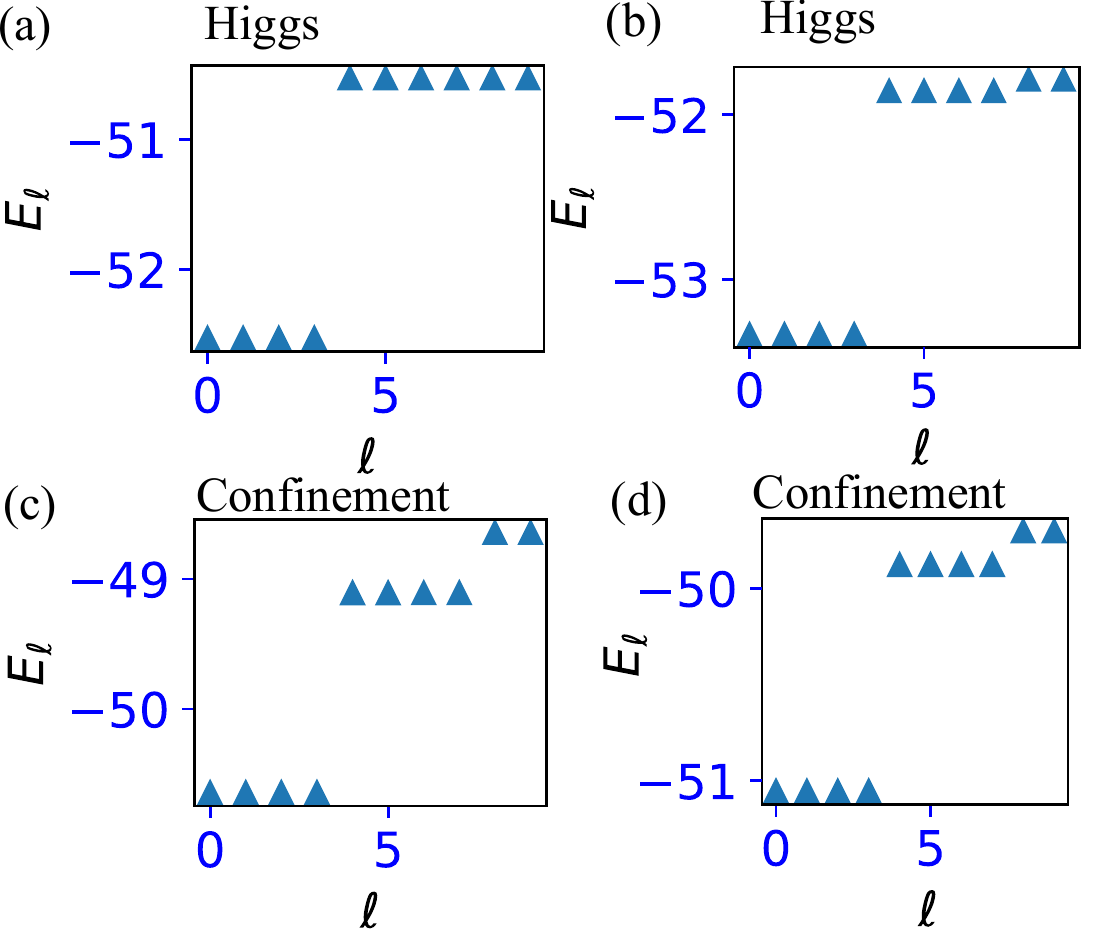}  
\end{center} 
\caption{(a) and (b): Low-lying energy spectra for the deep Higgs phase.
(a) $J^x_{\ell}=0$ ($W_\gamma$-symmetry exact), (b) $J^x_{\ell}=-0.5$ (explicit breaking of $W_\gamma$-symmetry).
For both cases, $h_v=-1$, $h_p=-1$, $J^z_{\ell}=-4$.
(c) and (d): Low-lying energy spectra for the deep confinement phase.
(c) $J^z_{\ell}=0$ ($H_{\gamma}$-symmetry exact), (d) $J^z_{\ell}=-0.5$ (explicit breaking of $H_\gamma$-symmetry).
For both cases, $h_v=-1$, $h_p=-1$, $J^x_{\ell}=-4$. 
The setup of the system for calculation is shown in Fig.~\ref{Fig3}(a), in which the total number of link is 18, and $(L_x,L_y)=(2,4)$.}
\label{Fig4}
\end{figure}

Before going to show detailed numerics, we briefly verify this expectation for $J^z_\ell=-4$, where the system for numerical calculation is shown in Fig.~\ref{Fig3}(a). 
The results are shown in Figs.~\ref{Fig4}(a) and (b), and we observe four-fold degeneracy in ground states. 
Also  even for a small but finite value of $J^x_\ell=-0.5$, the four-fold degeneracy maintains.
As effects of the one-form symmetry $W_\gamma$ is robust, it protects the bulk LRO as long as the state is in the Higgs phase. 

The above result gives very important observation, i.e., the charge symmetry $P=S_X$ is
expected to exhibit SSB in the thermodynamic limit. In ordinary systems of gauge theory with periodic boundary conditions, 
condensation of charged operators is masked by the requirement of the local-gauge invariance. Order parameters, 
which are expected to signal the SSB of the charge symmetry, are non-local and charge-neutral objects such as $\langle \phi_x \exp(i\int_\Gamma A_\mu dx_\mu) \phi^\dagger_y\rangle$ with a charged matter field $\phi$, vector potential $A_\mu$ and line $\Gamma$ connecting $\phi_x$ and $\phi^\dagger_y$~\cite{Kennedy1985}.
In the present case, however, the symmetry operator $S_X$ is nothing but the charge operator $P$ by Gauss law, and the SSB of $S_X$, $\langle S_X \rangle =\langle P\rangle= 0$, strictly indicates the SSB of the global charge.

In order to avoid confusion and misunderstanding, here we emphasize that the specific boundary conditions do \textit{not} induce the SSB but \textit{triggers} the boundary SSB, 
which distinguishes Higgs and confinement phases.
A suitable order parameter and its conjugate, $\tilde{Z}$ and $\tilde{X}$,  are supplied by the boundary conditions,
and they clearly describe the boundary SSB~\cite{note2}.
Monte-Carlo simulation of  the system $H_{\rm GHM}$ with the rough-smooth boundary conditions is interesting 
and desired to see if bulk or 1D critical behavior emerges in the specific heat, etc.
This is a future problem to be worked without difficulties.

Similar argument can be applied to the confinement of charged particles because of the electric-magnetic 
duality.
For $-J^x_\ell \gg |J^z_\ell|, |h_v|, |h_p|$, the LRO of $\sigma^x_\ell$ emerges with finite expectation value of order parameter
$\langle \tilde{X}\rangle\neq 0$ on the smooth boundaries indicating the SSB of the $\tilde{Z}$-symmetry, $\langle Z\rangle =0$, in the thermodynamic limit.
This means the emergence of condensation of magnetic charge (magnetic flux), which causes strong fluctuations of the gauge field $\sigma^z_\ell$, and charge confinement. 
Numerical calculations in Figs.~\ref{Fig4}(c) and ~\ref{Fig4}(d) clearly show the four-fold degeneracy of the ground state in confinement regime, hence supporting the above consideration for both Higgs and confinement phases.
Existence of the one-form symmetry $W_\gamma(H_\gamma)$ makes this state robust against 
the $J^x_\ell(J^z_\ell)$-terms in the Hamiltonian.
The above qualitative picture of the Higgs mechanism and quark confinement is not new, but the present model explicitly shows these mechanisms without any obscurity.

In the following subsections, we shall consider the model in a finite system from the viewpoint of subsystem quantum code and logical qubit. 
We shall derive the analytical description of the encoded qubit states, and introduce notion of strong zero modes, which are  generated by $(\tilde{X},\tilde{Z})$ and $(\tilde{X}^\ast,\tilde{Z}^\ast)$, and play an important role in the 
present work.

\subsection{Higgs phase}
In this subsection, we consider the deep Higgs regime such as $J^z_{\ell}\to -\mbox{large}$ , $h_v=-1$, $h_p=-1$, and $|J^{x}_\ell|\ll |J^z_{\ell}|$. 
The bulk LRO $\sigma^z_\ell \sim 1$ induces an effective Hamiltonian of the boundary spins similar 
to the transverse field Ising model (TFIM) as mentioned in \cite{Verresen2022}.

For both top and bottom rough boundaries, the $P$ symmetry is spontaneously broken in the Higgs phase. 
Then on the edges, cat states emerge in a finite system, and therefore, the boundary states are given by 
\begin{eqnarray}
|\pm\rangle_{\rm T(B)}\equiv \frac{1}{\sqrt{2}}
[|\Uparrow\rangle_{\rm T(B)}\pm |\Downarrow\rangle_{\rm T(B)}],
\end{eqnarray}
where ${\rm T(B)}$ stands for the top (bottom) rough boundary, $|\Uparrow\rangle_{\rm T(B)}$ and $|\Downarrow\rangle_{\rm T(B)}$ are all spin up 
and down states on dangling links on each rough boundary, $|\Uparrow\rangle =|\uparrow\uparrow\cdots \uparrow\rangle$ and $|\Downarrow\rangle =|\downarrow\downarrow\cdots \downarrow\rangle$. 
The boundary states $|\pm\rangle_{\rm T(B)}$ is $Z_2$ cat states, the presence of which is justified by SSB of $P$ symmetry 
on the top and bottom boundaries 

In the Higgs phase, each edge of the system induces two-fold degeneracy, thus, we expect the ground state of the Higgs phase is four-fold degenerate. 
The four degenerate states for the whole system are given 
as follows in the deep Higgs limit,
\begin{eqnarray}
&&|G1\rangle\equiv |+\rangle_{\rm T}|-\rangle_{\rm B}\otimes|\mbox {bulk}\rangle,\nonumber\\
&&|G2\rangle\equiv |-\rangle_{\rm T}|+\rangle_{\rm B}\otimes|\mbox{ bulk}\rangle,\nonumber\\
&&|G3\rangle\equiv |+\rangle_{\rm T}|+\rangle_{\rm B}\otimes|\mbox{ bulk}\rangle,\nonumber\\
&&|G4\rangle\equiv |-\rangle_{\rm T}|-\rangle_{\rm B}\otimes|\mbox{ bulk}\rangle, 
\label{1qubit_Higgs}
\end{eqnarray}
where $|{\rm bulk}\rangle$ represents the bulk states not including links on rough boundaries.
The above explicit form of the four-fold degenerate ground state for $H_{{\rm TC}}$ is useful for investigation by
numerical methods given later on.
Actually, we identify two pairs in a  topologically-symmetric sector. 
That is, for the sector $(P,S_{z})=(-1,+1)$, the pair of the ground state is $\{ |G1\rangle,|G2\rangle \}$ 
since $P|G1(2)\rangle=(-1)|G1(2)\rangle$ and $S_Z|G1(2)\rangle=(+1)|G1(2)\rangle$. 
This pair is an encoded qubit 
satisfying $\tilde{Z}|G1(2)\rangle=|G2(1)\rangle$, $\tilde{X}|G1\rangle=(+1)|G1\rangle$ and $\tilde{X}|G2\rangle=(-1)|G2\rangle$. 
Note that making the boundary cat state is an essential ingredient to get a qubit controlled by
the logical operators $(\tilde{X},\tilde{Z})$.

The other pair $\{ |G3\rangle,|G4\rangle \}$ in the four-fold degenerate ground state is another encoded qubit. 
This pair belongs to the sector $(P,S_{z})=(+1,+1)$ since $P|G3(4)\rangle=(+1)|G3(4)\rangle$ and $S_Z|G3(4)\rangle=(+1)|G3(4)\rangle$. 
They also satisfy $\tilde{Z}|G3(4)\rangle=|G4(3)\rangle$ and $\tilde{X}|G3\rangle=(+1)|G3\rangle$ and $\tilde{X}|G4\rangle=(-1)|G4\rangle$. 

Due to the robustness of the one-form symmetry, the bulk LRO survives even in 
the presence of its explicit breaking terms in the Hamiltonian.
As a result, the above-encoded qubit with cat-state properties is preserved intact. 
We shall demonstrate this observation by numerical methods later on.

\subsection{Confinement phase}
We next consider the deep confinement limit, $J^x_{\ell}\to -\mbox{large}$, 
$h_v=-1$, $h_p=-1$, and $|J^{z}_\ell|\ll |J^x_{\ell}|$. 
The bulk LRO $\sigma^x_\ell \sim 1$ induces an effective Hamiltonian for the smooth-boundary spins similar
to TFIM.

For both left and right smooth boundaries, the $S_Z$ symmetry is spontaneously broken in the confinement phase. 
Then on the smooth boundaries, the boundary states are given by, 
\begin{eqnarray}
|\pm_x\rangle_{\rm L(R)}\equiv \frac{1}{\sqrt{2}}[|\Rightarrow\rangle_{\rm L(R)}\pm |\Leftarrow\rangle_{\rm L(R)}],
\end{eqnarray}
where the subscript ${\rm L(R)}$ stands for the left (right) smooth boundary, $|\Rightarrow\rangle_{\rm L(R)}$ and $|\Leftarrow\rangle_{\rm L(R)}$ are all positive $x$ and negative $x$ spin states on each smooth boundary, respectively, i.e., $|\Rightarrow\rangle =|\rightarrow\rightarrow\cdots \rightarrow\rangle$ and $|\Leftarrow\rangle =|\leftarrow\leftarrow\cdots \leftarrow\rangle$. 
The boundary states $|\pm_{x}\rangle_{\rm L(R)}$ are $Z_2$-cat states composed of $|\Rightarrow\rangle$ and $|\Leftarrow\rangle$, the emergence of which 
is induced by SSB of $S_Z$ symmetry at the left and right boundaries. 

We expect that the ground-state structure is analogue to that of the Higgs phase. 
That is, each edge of the system induces two-fold degeneracy and therefore, the ground state is four-fold degenerate. 
Furthermore, the four-fold degenerate states for the whole system are given by
\begin{eqnarray}
&&|C1\rangle\equiv |+_x\rangle_{\rm L}|-_{x}\rangle_{\rm R}\otimes|\mbox {bulk}\rangle,\nonumber\\
&&|C2\rangle\equiv |-_x\rangle_{\rm L}|+_x\rangle_{\rm R}\otimes|\mbox{ bulk}\rangle,\nonumber\\
&&|C3\rangle\equiv |+_x\rangle_{\rm L}|+_x\rangle_{\rm R}\otimes|\mbox{ bulk}\rangle,\nonumber\\
&&|C4\rangle\equiv |-_x\rangle_{\rm L}|-_x\rangle_{\rm R}\otimes|\mbox{ bulk}\rangle.
\label{1qubit_confine}
\end{eqnarray}
The above explicit form of the four-fold degenerate ground state for $H_{{\rm TC}}$ in the confinement limit is useful for later investigation by the numerical methods.
Here, we should mention that the ground states in Eqs.~(\ref{1qubit_Higgs}) and (\ref{1qubit_confine}) exhibit
a typical form of the subsystem code considered in Ref.~\cite{Wildeboer2022}. 

The four states in Eq.~(\ref{1qubit_confine}) again behave as two pairs of qubit.
For the sector $(P,S_{z})=(+1,-1)$, the relevant pair of the ground state is $\{ |C1\rangle,|C2\rangle \}$ 
since $P|C1(2)\rangle=(+1)|C1(2)\rangle$ and $S_Z|C1(2)\rangle=(+1)|C1(2)\rangle$. 
These paired states form an encoded qubit, that is, 
$\tilde{X}|C1(2)\rangle=|C2(1)\rangle$ and $\tilde{Z}|C1\rangle=+1|C1\rangle$ and $\tilde{Z}|C2\rangle=-1|C2\rangle$. 
Similarly, the other paired states $\{ |C3\rangle,|C4\rangle \}$ also form an encoded qubit  in the sector 
$(P,S_{z})=(+1,+1)$. 
The logical operators are similarly given by $\tilde{Z}$ and $\tilde{X}$. 

Here, we should note that the states in Eq.~(\ref{1qubit_confine}) are {\it exactly dual} to those of the Higgs phase in Eq.~(\ref{1qubit_Higgs})  under the interchange  $\sigma^x_{\ell} \Longleftrightarrow \sigma^z_{\ell}$ for all links. 
However sometimes, this duality is broken explicitly by the geometry of the lattice if $L_x\neq L_y$, where $L_{x(y)}$ is the number of links 
in the horizontal (vertical) direction, as the rectangular lattice used
in our numerical studies.  (See Fig.~\ref{Fig3}.)

We remark that the wavefunctions in Eqs.~(\ref{1qubit_Higgs}) and (\ref{1qubit_confine}) have a standard form of the subsystem code such as $|\psi_L\rangle|\psi'\rangle$~\cite{Wildeboer2022}, where $|\psi_L\rangle$ 
is the state of the logical qubit and $|\psi'\rangle$ is that of the gauge qubit. In the Higgs and confinement regimes, the states $|\psi_L\rangle$ 
can be explicitly obtained in a compact form, whereas it cannot in the deconfinement regime. 
This is also the case for more general cases of multiply-encoded qubits and higher-energy qubits, which we discuss in the subsequent subsections.
Therefore, on constructing subsystem codes in the present model, the Higgs and confinement regimes are better than the deconfinement regime. 
The practical representation of the wavefunctions obtained in the above is useful for investigating stability and error corrections of 
the subsystem codes and also may give an insight into the design of quantum memories robust to decoherence, which are future problems, 
although we will briefly mention the fault tolerance of the present subsystem codes to thermal noise \cite{ex_Nussinov2008}.

\subsection{Multiply-encoded qubits}
By manipulating the shape of the boundary, multiply-encoded qubit can be 
constructed~\cite{Wildeboer2022}. 
For example, a doubly-encoded qubit subsystem code can be put on three rough and smooth boundaries as shown in Fig.~\ref{Fig3}(b). 
In the Higgs and confinement phases, we can explicitly describe the state of these qubits using 
ideas of the gauge theory. 
In this subsection, we mostly focus on the deep Higgs regime $J^{z}_{\ell}\to -$large, since duality
between Higgs and confinement phases exists even in this multiply-encoded qubits system.

\underline{Doubly-encoded qubits}:
As we mentioned in the above,
we employ the system displayed in Fig.~\ref{Fig3}(b).
The ground state of $H_{\rm TC}$ for the Higgs and confinement phase is eight-fold degenerate. 
Here, these states are decomposed into two distinct $(P,S_Z)$ sectors. Each sector includes four-fold degenerate eigenstates.

In the deep Higgs phase, we find that one set of four-fold degenerate ground states are given as
\begin{eqnarray}
&&|Q2_1\rangle\equiv |+\rangle_{r1}|+\rangle_{r2}|+\rangle_{r3}\otimes|\mbox {bulk}\rangle,\nonumber\\
&&|Q2_2\rangle\equiv |+\rangle_{r1}|-\rangle_{r2}|-\rangle_{r3}\otimes|\mbox{ bulk}\rangle,\nonumber\\
&&|Q2_3\rangle\equiv |-\rangle_{r1}|+\rangle_{r2}|-\rangle_{r3}\otimes|\mbox{ bulk}\rangle,\nonumber\\
&&|Q2_4\rangle\equiv |-\rangle_{r1}|-\rangle_{r2}|+\rangle_{r3}\otimes|\mbox{ bulk}\rangle,
\end{eqnarray}
where $r_{\ell}$ ($\ell=1,2,3$) represents three rough boundaries in Fig.~\ref{Fig3}(b) and $|{\rm bulk}\rangle$ represents the bulk state without including boundary links. 
Then, for these four states, two set of logical operators exist \cite{Wildeboer2022}, which are given by
\begin{eqnarray}
\tilde{X}_1&=&\prod_{k\in s1}\sigma^z_k,\:\:
\tilde{X}_2=\prod_{k\in s2}\sigma^z_{k},\nonumber\\
\tilde{Z}_1&=&\prod_{m\in r1}\sigma^x_m,\:\:
\tilde{Z}_2=\tilde{Z}_{1}\prod_{m\in r2}\sigma^z_m.
\label{logical_2Q}
\end{eqnarray}
These operator act on the above four-fold degenerate ground states suitably, and the states work as a doubly-encoded qubit.

The other four-fold degenerate ground states also work as a doubly-encoded qubit, and they are
explicitly given as
\begin{eqnarray}
&&|Q2_5\rangle\equiv |+\rangle_{r1}|+\rangle_{r2}|-\rangle_{r3}\otimes|\mbox {bulk}\rangle,\nonumber\\
&&|Q2_6\rangle\equiv |+\rangle_{r1}|-\rangle_{r2}|+\rangle_{r3}\otimes|\mbox{ bulk}\rangle,\nonumber\\
&&|Q2_7\rangle\equiv |-\rangle_{r1}|+\rangle_{r2}|+\rangle_{r3}\otimes|\mbox{ bulk}\rangle,\nonumber\\
&&|Q2_8\rangle\equiv |-\rangle_{r1}|-\rangle_{r2}|-\rangle_{r3}\otimes|\mbox{ bulk}\rangle.
\end{eqnarray}
The operators of Eq.~(\ref{logical_2Q})  again act on the above states as the logical operators.

In the deep confinement phase, the states dual for the above degenerate ground states are also regarded as doubly-encoded qubits, 
and there the logical operators are interchanged with each other, $\tilde{X}_{\ell} \longleftrightarrow \tilde{Z}_{\ell}$ $(\ell=1,2)$.\\

\underline{General $N$-encoded qubits}:
General $N$-encoded qubits can be designed in the system with many rough and smooth boundaries~\cite{Wildeboer2022}. 
One of lattices for that system is displayed in Fig.~\ref{Fig3}(c), where to implement $N$-encoded qubits, $N+1$ rough and smooth boundaries are needed and their locations 
must be alternate.
As in the previous cases, we can give an explicit analytical description of the $N$-encoded qubit states with the boundary SSB in the deep Higgs and confinement phases. 
For the system $H_{\rm TC}$, we expect that the ground state is $(2\times 2^N)$-fold degenerate in the deep Higgs phase. 
Then, the first set of $2^N$-degenerate ground states of $N$-encoded qubits are given by
\begin{eqnarray}
|z_1,z_2,\cdots, z_{N}\rangle \equiv \biggl[\bigotimes^{N}_{\ell=1}|{z_{\ell}}\rangle\biggr]|z_{N+1}\rangle_{r_{\ell}}\otimes |\mbox{bulk}\rangle,
\label{Nqubit}
\end{eqnarray}
where $z_{\ell}=\pm$, $|z_{\pm}\rangle=|\pm\rangle_{r_{\ell}}$, $r_\ell$ relabels the $r_{\ell}$-th rough boundary as shown in Fig.~\ref{Fig3}(c) and 
\begin{eqnarray}
z_{N+1}=(-1)^{N+\sum^{N}_{\ell=1}(z_{\ell}+1)/2}.
\end{eqnarray}
The label $z_{\ell}$ ($\ell=1,2,\cdots,N$) denotes up or down state of the $\ell$-encoded qubits. 
The above set of $N$-qubits is embedded in  $(P,S_Z=+1)$ sector, where $P=\biggr[\prod^{N}_{1}z_{\ell}\biggr]z_{N+1}\equiv P_N$.

The second set of $2^N$-degenerate ground states of $N$-encoded qubits in ground-state multiplet are similarly  given by Eq.~(\ref{Nqubit}) with replacing as
\begin{eqnarray}
z_{N+1}\to z_{N+1}=(-1)^{N+1+\sum^{N}_{\ell=1}(z_{\ell}+1)/2}.
\end{eqnarray}
The $2^N$-degenerate ground states are embedded in the sector $(P,S_Z)=(-P_N,+1)$.

For these $N$-encoded qubits, the logical operators are given as \cite{Wildeboer2022}
\begin{eqnarray}
\tilde{X}_n=\prod_{k\in r_{n}}\sigma^z_k,\:\:
\tilde{Z}_n=\prod^{n}_{m=1}\tilde{Z}^s_m,
\label{logical_NQ}
\end{eqnarray}
where $n$ denotes the number of encoded qubit taking $n=1,2,\cdots, N$,
$r_n$ is $n$-th rough boundary shown in Fig.~\ref{Fig3}(c), $n=1,2,\cdots, N$, and
\begin{eqnarray}
\tilde{Z}^s_m=\prod_{q\in s_m}\sigma^z_q.
\end{eqnarray}
Here $s_m$ is $m$-th smooth boundary shown in Fig.~\ref{Fig3}(c), $m=1,2,\cdots, N$.

As in the previous cases, the dual of the above qubits states are obtained straightforwardly to get
$N$-qubit states in the deep confinement phase. Also, the dual of the logical operators of Eq.~(\ref{logical_NQ}) act as logical operators on the dual states.

\subsection{Excited states and single-encoded qubits in Higgs and confinement phases}

The previous work \cite{Wildeboer2022}
suggested that the encoded qubit can be embedded in arbitrary excited states, as natural properties of the subsystem code \cite{Poulin2005}.
In this subsection, we study an explicit analytical description of the encoded qubits in excited states. 
These properties of excited states are closely related to the strong zero mode \cite{Fendley2012,Fendley2016}, which has been discussed in various contexts.

As shown in the previous section, the states of the qubit in both deep Higgs and confinement phase are given via the bulk-boundary factorised form such as 
\begin{eqnarray}
|Q\rangle = | \mbox{boundary} \rangle \otimes |\mbox{bulk}\rangle,
\end{eqnarray}
where the bulk is a simple product state and the boundary is a cat state.
Here, we assume that the excitation energy of the bulk $\Delta E_{\rm bulk}$ is much larger than that of 
the boundary $\Delta E_{\rm b}$, $\Delta E_{\rm bulk}\gg \Delta E_{\rm b}$, which is satisfied for $|J^{x(z)}_{\ell}| \gg |h_{v}|, |h_p|, |J^{z(x)}_{\ell}|$. 
Then, low-energy excited states of the encoded qubit are constructed only by the boundary excitation in $|\mbox{boundary}\rangle$. 
For example in the deep Higgs phase, the boundary states are cat states, and therefore,
the excited states are constructed upon the cat state residing on one of the rough boundaries \cite{Kemp2017}, 
which are explicitly given by 
\begin{eqnarray}
|e^k_{\pm}\rangle \equiv \frac{1\pm P}{\sqrt{2}}|\uparrow\rangle_{j_0}\otimes|\{\uparrow\}_k \{\downarrow\}_{\ell_e-k-1}\rangle_{\nin j_0},
\label{e_cat}
\end{eqnarray}
where $j_0$ is a link at one end of a rough boundary, $\ell_{e}$ is the number of the link of one 
rough boundary, and $0\le k\le\ell_e-1$. 
The state $|\{\uparrow\}_k\{\downarrow\}_{\ell_e-k-1}\rangle_{\nin j_0}$ is a product state composed of $k$ up-spins and $(\ell_e-k-1)$ down-spins, 
in which location of the domain wall is arbitrary. 
The low-energy excited states without bulk excitation can be obtained by replacing the cat state $|\pm\rangle$ in states of Eq.~(\ref{1qubit_Higgs}) 
with the state $|e^k_{\pm}\rangle$ in Eq.~(\ref{e_cat}). 
As an example, excited encoded qubit states constructed from the states $|G1\rangle$-$|G4\rangle$ in 
Eq.~(\ref{1qubit_Higgs}) are given by 
\begin{eqnarray}
&&|G1e^k\rangle\equiv |e^k_{+}\rangle_{\rm T}|e^k_{-}\rangle_{\rm B}\otimes|\mbox {bulk}\rangle,\nonumber\\
&&|G2e^k\rangle\equiv |e^k_{-}\rangle_{\rm T}|e^k_{+}\rangle_{\rm B}\otimes|\mbox{ bulk}\rangle,\nonumber\\
&&|G3e^k\rangle\equiv |e^k_{+}\rangle_{\rm T}|e^k_{+}\rangle_{\rm B}\otimes|\mbox{ bulk}\rangle,\nonumber\\
&&|G4e^k\rangle\equiv |e^k_{-}\rangle_{\rm T}|e^k_{-}\rangle_{\rm B}\otimes|\mbox{ bulk}\rangle.
\label{1qubit_Higgs_ex}
\end{eqnarray}
Here, pairs $(\{ |G1e^k\rangle,|G2e^k\rangle \})$ and 
$(\{ |G3e^k\rangle,|G4e^k\rangle \})$ are single-encoded qubits embedded into excited states in the deep Higgs phase. 
By this manipulation, various types of degenerate eigenstates of encoded qubit can be constructed, and also
by using duality, the low-energy excited states constructing an encoded qubit are obtained for the deep confinement phase.

Furthermore, general higher-excited states are to be described in the following way. 
In the case in which pure bulk excitations emerge in a plaquette or at a vertex not-touching the boundaries, the excited states can be written 
as a product state composed of cat states on the boundaries and bulk excited states. 
As a whole, we expect that most of the states having properties of encoded qubit are to be described in the Higgs and confinement phases.\\

\subsection{Entangled encoded qubit states in $H_{\rm GHM}$}
In the previous subsections, we showed the explicit form of the logical encoded qubit states for the disentangled Hamiltonian, $H_{{\rm TC}}$. 
In general, the encoded qubit state denoted by $|Q_{{\rm TC}}\rangle$ can be transformed into the encoded qubit state in the original (2+1)-D $Z_2$ extended gauge-Higgs model, 
$H_{\rm GHM}$, denoted by $|Q_{\rm GHM}\rangle$ as 
\begin{eqnarray}
|Q_{\rm GHM}\rangle = (U_vU_p)^\dagger|Q_{{\rm TC}}\rangle=U_pU_v|Q_{{\rm TC}}\rangle,
\end{eqnarray}
since undoing the disentangling is manipulated by using Gauss-law constraints, and states are
uniquely determined up to the local-gauge symmetries.
Under this transformation, the boundary state and bulk state in the state $|Q_{{\rm TC}}\rangle$ become moderately entangled. 
In particular, we expect that the Higgs phase regime, the encoded qubit state $|Q_{\rm GHM}\rangle$ in the Higgs ground state multiplet can be short-range entangled state 
since the state is expected to be the SPT (2D cluster state) \cite{Borla2021,Verresen2022}, also due to duality, the confinement is the same.
For example, spin operators in dangling links of the rough boundaries in $H_{\rm TC}$ correspond to
gauge-invariant operators in $H_{\rm GHM}$ via $\sigma^z_{\ell_v} \Longleftrightarrow \sigma^z_{\ell_v} Z_v$.
This means, e.g.,  $|+\rangle_{\rm TC} \Longleftrightarrow {1 \over \sqrt{2}}
(|+,+\rangle_{\rm GHM}+|-,-\rangle_{\rm GHM})$,
where notations are self-evident.
In such a way, short-range entanglement emerges between the gauge field and matter fields. \\

\begin{figure}[t]
\begin{center} 
\includegraphics[width=8cm]{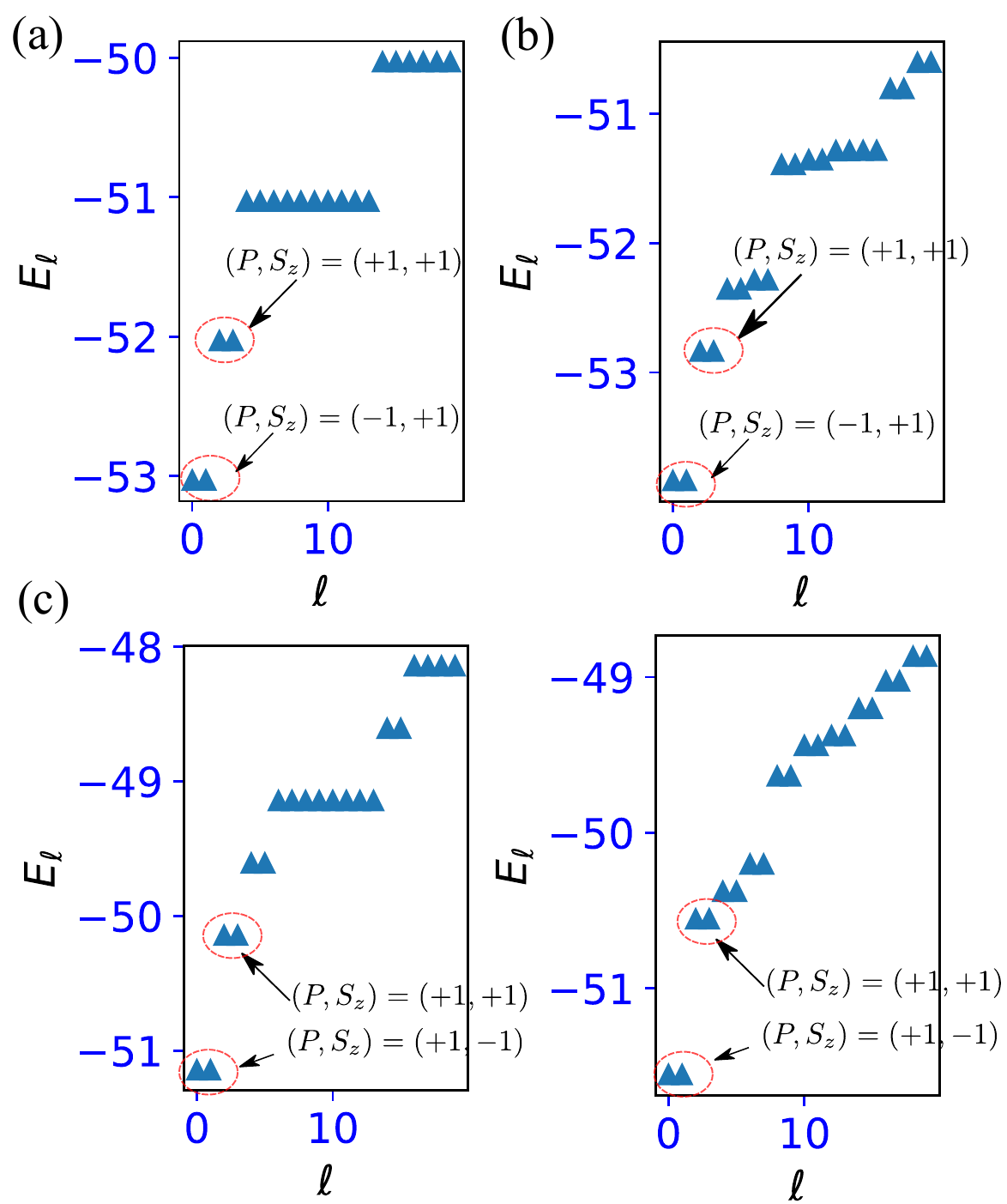}  
\end{center} 
\caption{Signature of doubly-encoded qubit. 
(a) and (b): Low-lying energy spectra for the deep Higgs phase. (a) $J^x_{\ell}=0$($W_\gamma$-symmetry exact), (b) $J^x_{\ell}=-0.5$ (explicit breaking of $W_{\gamma}$-symmetry).
For both cases, $h_v=-1$, $h_p=-1$, $J^z_{\ell}=-4$, $v_1=0.5$, $v_2=0$.
The first and the second pair energies are splitted $\sim 2v_1$.
(c) and (d): Low-lying energy spectra for the deep confinement phase.
(c) $J^z_{\ell}=0$ ($H_\gamma$-symmetry exact), (d) $J^z_{\ell}=-0.5$ (explicit breaking of $H_\gamma$-symmetry). 
For both cases, $h_v=-1$, $h_p=-1$, $J^x_{\ell}=-4$, $v_1=0$, $v_2=0.5$.
The first pair and the second pair energies split with a gap $\sim 2v_2$.
}
\label{Fig5}
\end{figure}

\section{Numerical study for small systems}
In this section, we investigate the ground state properties of the disentangle Hamiltonian in detail by numerical methods. 
In particular, to identify the degeneracy of the encoded qubits of the subsystem code, we add the potential $V_{\rm pot}=v_1 P+v_2 S_Z$ to the Hamiltonian $H_{\rm TC}$, 
and then, the system is described by $H_{{\rm TC}}+V_{\rm pot}$.

\subsection{Subsystem code of single-encoded qubit in low-energy spectrum}
We diagonalize the Hamiltonian $H_{{\rm TC}}+V_{\rm pot}$ by employing Quspin package \cite{Quspin1, Quspin2} and calculate low-energy spectrum up to $20$-$30$ th state 
from the ground state. 
In what follows, we set $h_v=h_p=-1$ and focus on both the deep Higgs and confinement phases. 
The lattice structure is shown in Fig.~\ref{Fig3}(a), where each rough and smooth boundaries have three and four vertical links, respectively, 
and the total number of  links is eighteen, i.e., $(L_x,L_y)=(2,4)$.

\underline{Deep Higgs phase}:
We first study the deep Higgs regime and set the parameters as $J^z_{\ell}=-4$ and $(v_1,v_2)=(0.5,0)$. 
The obtained spectrum for the $W_{\gamma}$-symmetric case $J^x_\ell=0$ is shown in Fig.~\ref{Fig5}(a). 
We observe that the four states in the degenerate ground-state multiplet for $(v_1,v_2)=(0,0)$ 
split into two pairs, each of which is nothing but single-encoded qubits in the sector $(P,S_Z)=(+1,-1)$ and $(+1,+1)$, respectively. 
The above result agrees with the analytical study in the previous section, in particular, the emergent sectors. 
We observe that the higher-energy states belong to a degenerate multiplet with more than two even-number states. 
This degeneracy of the energy spectrum is stable against the explicit breaking of the $W$-symmetry
by the $J^x_{\ell}$-term in the Hamiltonian, as the results for $J^x_{\ell}=-0.5$ in Fig.~\ref{Fig5}(b) indicate
that the ground state and first excited pairs are single-encoded qubits in the sector $(P,S_Z)=(+1,-1)$ and $(+1,+1)$, respectively. 
Also for finite values of $J^x_{\ell}$, the higher-excited states tend to be two-fold degenerate, i.e.,  
the splitting into two pairs is enhanced by the $J^x_{\ell}$-term.

\begin{figure}[t]
\begin{center} 
\includegraphics[width=8cm]{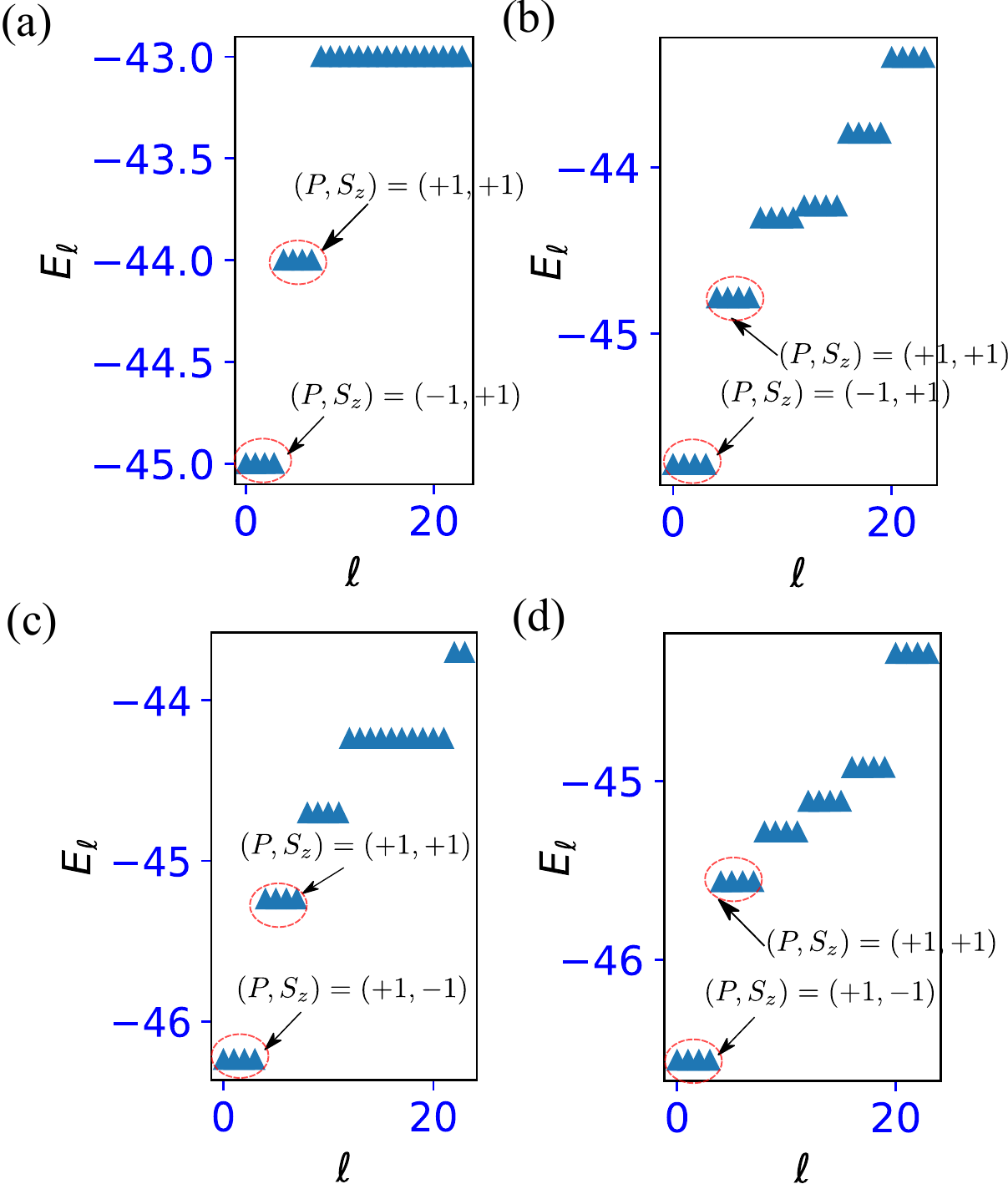}  
\end{center} 
\caption{Signature of doubly-encoded qubits: 
Low-lying energy spectra for the deep Higgs phase. (a) $J^x_{\ell}=0$ ($W$-symmetry exact), (b) $J^x_{\ell}=-0.5$ (explicit breaking of $W_\gamma$-symmetry)
For both cases, $h_v=-1$, $h_p=-1$, $J^z_{\ell}=-4$, $v_1=0.5$, $v_2=0$. 
The first and second four-pair energies are splitted $\sim 2v_1$.
(c) and (d): Low-lying energy spectra for the deep confinement phase.
(c) $J^z_{\ell}=0$ ($H_\gamma$-symmetry exact), (d) $J^z_{\ell}=-0.5$ (explicit breaking of $H_\gamma$-symmetry)
For both cases, $h_v=-1$, $h_p=-1$, $J^x_{\ell}=-4$, $v_1=0$, $v_2=0.5$.
The first and second four-pair energies split with a gap $\sim 2v_2$.
}
\label{Fig6}
\end{figure}
\underline{Deep confinement phase}:
We turn to the deep confinement regime, and put $J^x_{\ell}=-4$ and set $(v_1,v_2)=(0,0.5)$ to examine duality between Higgs and confinement regimes. 
The spectrum for the $H_\gamma$-symmetric case $J^x_\ell=0$ is shown in Fig.~\ref{Fig5}(c). 
We again observe that the four-fold degenerate ground-state multiplet for $(v_1,v_2)=(0,0)$ splits into two pairs, each of which  is a single-encoded qubit in the sector 
$(P,S_Z)=(-1,+1)$ and $(+1,+1)$, respectively. 
The above result agrees with the analytical observation in the previous section. 
The low-energy degenerate structure is stable against the explicit breaking of $H_{\gamma}$-symmetry by finite
values of $J^x_{\ell}$ as shown in Fig.~\ref{Fig5}(d). 
The spectrum is almost the same as that of the deep Higgs regime connected by duality,
with a small but finite discrepancy coming from $L_x\neq L_y$.

\subsection{Subsystem code of doubly-encoded qubit}
We next numerically observe doubly-encoded qubits existing in the system shown in 
Fig.~\ref{Fig3}(b), in which the total number of  link is eighteen.   
We employ a similar procedure, i.e., diagonalize the Hamiltonian $H_{{\rm TC}}+V_{\rm pot}$ to calculate the low-energy spectrum.

\underline{Deep Higgs phase}:
We set the parameters in the deep Higgs phase, $J^z_{\ell}=-4$ and also $(v_1,v_2)=(0.5,0)$. 
The spectrum for $W_{\gamma}$-symmetric case $J^x_\ell=0$ is shown in Fig.~\ref{Fig6}(a). 
We observe a four-fold degenerate ground state multiplet in the sector $(P,S_Z)=(-1,+1)$, which are nothing but two encoded qubits, and
the first excited four-fold degenerate states in the sector $(P,S_Z)=(+1,+1)$. 
The two four-fold degenerate states are predicted by the analytical study in the previous section. 

Similarly to the single-encoded qubit system, this low-energy degenerate structure is stable against the explicit breakdown of $W_{\gamma}$-symmetry by  
finite $J^x_{\ell}$ in the Hamiltonian. 
The result for the case with $J^x_{\ell}=-0.5$ is displayed in Fig.~\ref{Fig6}(b). 
We observe that the ground-state and first-excited  four-fold degenerate multiplets  are intact for finite $J^x$, 
and belong to the sector $(P,S_Z)=(-1,+1)$ and $(+1,+1)$, respectively. 
Also, for finite $J^x_{\ell}$, the higher-excited states tend to be four-fold degenerate, i.e., the four-fold degeneracy is enhanced.

\underline{Deep confinement phase}:
We turn to the deep confinement regime, where $J^x_{\ell}=-4$ and $(v_1,v_2)=(0,0.5)$. 
The spectrum for the $H_\gamma$-symmetric case $J^x_\ell=0$ is shown in Fig.~\ref{Fig6}(c). 
We again observe the four-fold degenerate ground state multiplet in the sector $(P,S_Z)=(+1,-1)$
as in the deep Higgs regime. 
These are two encoded qubit states. 
The first-excited four-fold degenerate states belong to the $(P,S_Z)=(+1,+1)$ sector. 
These results are in agreement with the analytical observation in the previous section. 
This degeneracy structure at low energies is intact for finite $J^x_{\ell}$ as shown in Fig.~\ref{Fig6}(d). 
Again for the doubly-encoded qubit system, the spectrum is almost the same as that
of the deep Higgs regime by duality.

\begin{figure}[t]
\begin{center} 
\includegraphics[width=8cm]{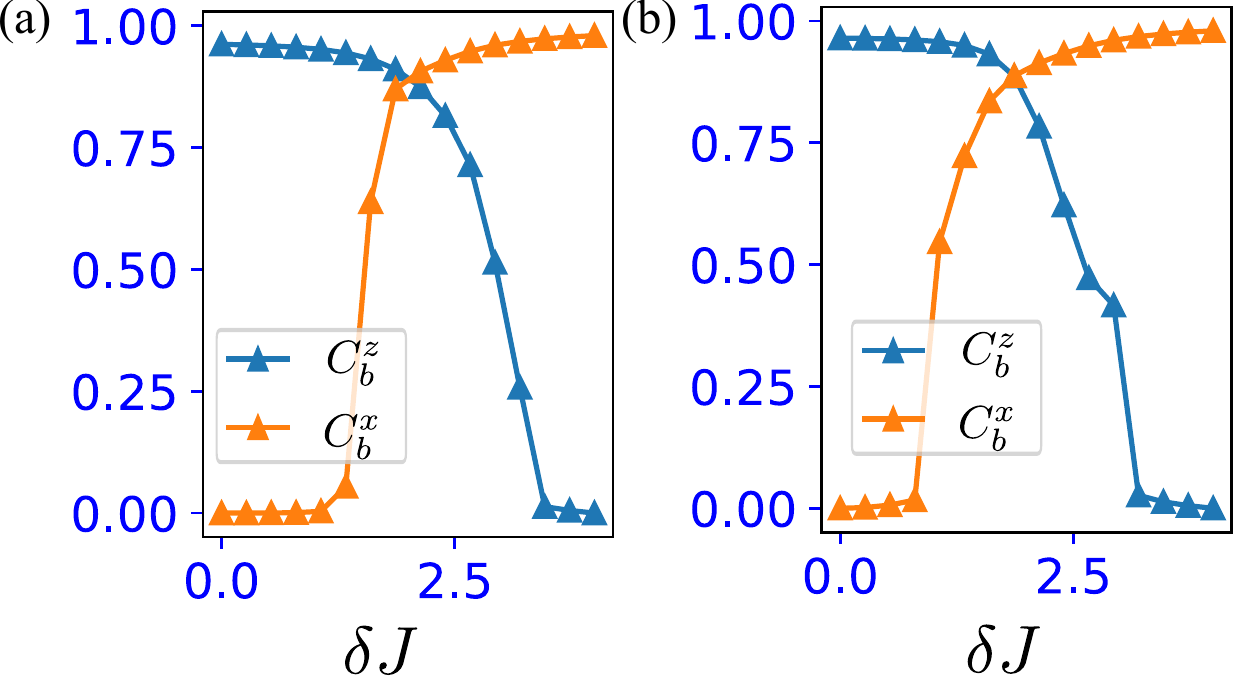} 
\end{center} 
\caption{Correlation functions $C^x_{\rm b}$ and $C^z_{\rm b}$ on boundaries for single-encoded qubit system [(a)] and doubly-encoded qubit system [(b)]. 
We set $(v_1,v_2)=(0.5,0)$ and $(0,0.5)$ for the calculations, $C^z_{\rm b}$ and $C^x_{\rm b}$.
}
\label{Fig7}
\end{figure}
\subsection{Higgs-confinement phase transition}
We numerically observe boundary correlation functions to observe the $S_X$($S_Z$)-SSB phase transition. 
For the single-encoded qubit shown in Fig.~\ref{Fig3}(a), the correlation functions are given by $C^z_{\rm b}=\langle \sigma^z_0\sigma^z_3\rangle$ and 
$C^x_{\rm b}=\langle \sigma^x_0\sigma^x_{15}\rangle$. 
The labels of the link are shown in Fig.~\ref{Fig3}(a). 
We calculate $C^z_{\rm b}$ and $C^x_{\rm b}$ for the $\tilde{Z}=+1$ sector of the two-fold degenerate ground state, where $J^{x}_\ell$ and $J^{z}_\ell$ are parameterized as $J^{z}_\ell=4-\delta J$ and $J^{x}_\ell=\delta J$ with varying $\delta J$. 

Figure \ref{Fig7}(a) exhibits the behavior of the correlations, where we set $(v_1,v_2)=(0.5,0)$ and $(0,0.5)$. 
We can observe a phase-transition-like behavior even in the small system. 
In the deep Higgs phase (small $\delta J$) $C^z_{\rm b}$ has a large finite value, while in the deep confinement 
phase (large $\delta J$) $C^x_{\rm b}$ is finite. 
This numerical result implies the emergence of the SSB of $P(S_X)$ or $S_Z$ symmetry in the thermodynamic limit. 

Similar behavior is observed in the doubly-encoded qubit system shown in Fig.~\ref{Fig3}(b). 
We define the correlation function as $C^z_{\rm b}=\langle \sigma^z_0\sigma^z_3\rangle$ and $C^x_{\rm b}=\langle \sigma^x_0\sigma^x_{15}\rangle$, where the labels of 
the link are shown in Fig.~\ref{Fig3}(b), and calculate $C^z_{\rm b}$ and $C^x_{\rm b}$ for the $(\tilde{Z_1},\tilde{Z_2})=(+1,+1)$ sector of the four-fold degenerate ground state. 
We plot the result in Fig.~\ref{Fig7}(b) to confirm the emergence of the SSB of $P(S_X)$ or $S_Z$ symmetry. 

From the all data shown in Fig.~\ref{Fig7}, we conclude that the SSB phase transition on the boundary occurs at $J^{z}_{\ell}\sim J^{x}_{x}$. 
This result indicates that the criticality of the present system coincides with that of the effective TFIM 
emerging on the boundary. 
However, we must be careful in concluding the existence of the phase transition because the system size
is very small. 
More elaborated numerical approaches such as quantum Monte-Carlo simulation (employed to a similar model \cite{Tupitsyn2010}) or DMRG are desired as a future work.

\subsection{Autocorrelator}
We numerically observed the degenerate energy spectra for both the Higgs and confinement phases. 
The degeneracy coming from the nature of encoded qubits survives to highly-excited eigenstates. 
This behavior of the system seems to indicate the  emergence of a strong zero mode, which was first
proposed in \cite{Fendley2012,Fendley2016}. 

We shall numerically verify the existence of the strong zero mode by observing the whole energy
spectrum of the present model by following the previous numerical study \cite{Kemp2017}. 
In that work, to examine the strong zero mode, autocorrelator measuring the coherence of logical operators was investigated in detail. 
We employ the same methods and calculate the autocorrelator of the logical operator in the present model,
which is  given by \cite{Kemp2017}
\begin{eqnarray}
C^s_{\rm au}(t)&=&\langle \psi_{s}|\tilde{X}(t)\tilde{X}(0)|\psi_{s}\rangle\nonumber\\
&=&\sum_{\ell}|\langle \psi_s|\tilde{X}|\psi_{\ell}\rangle|^2e^{i(E_{\ell}-E_{s})t},
\end{eqnarray}
where $\tilde{X}(t)=e^{it(H_{{\rm TC}}+V_{\rm pot})}\tilde{X}e^{-it(H_{{\rm TC}}+V_{\rm pot})}$ (we have set $\hbar=1$), and $|\psi_s\rangle$ is $s$-th eigenstates of 
$H_{{\rm TC}}+V_{\rm pot}$, and $E_{\ell}$ is $\ell$-th energy for the ascending order.

\begin{figure}[t]
\begin{center} 
\includegraphics[width=8.5cm]{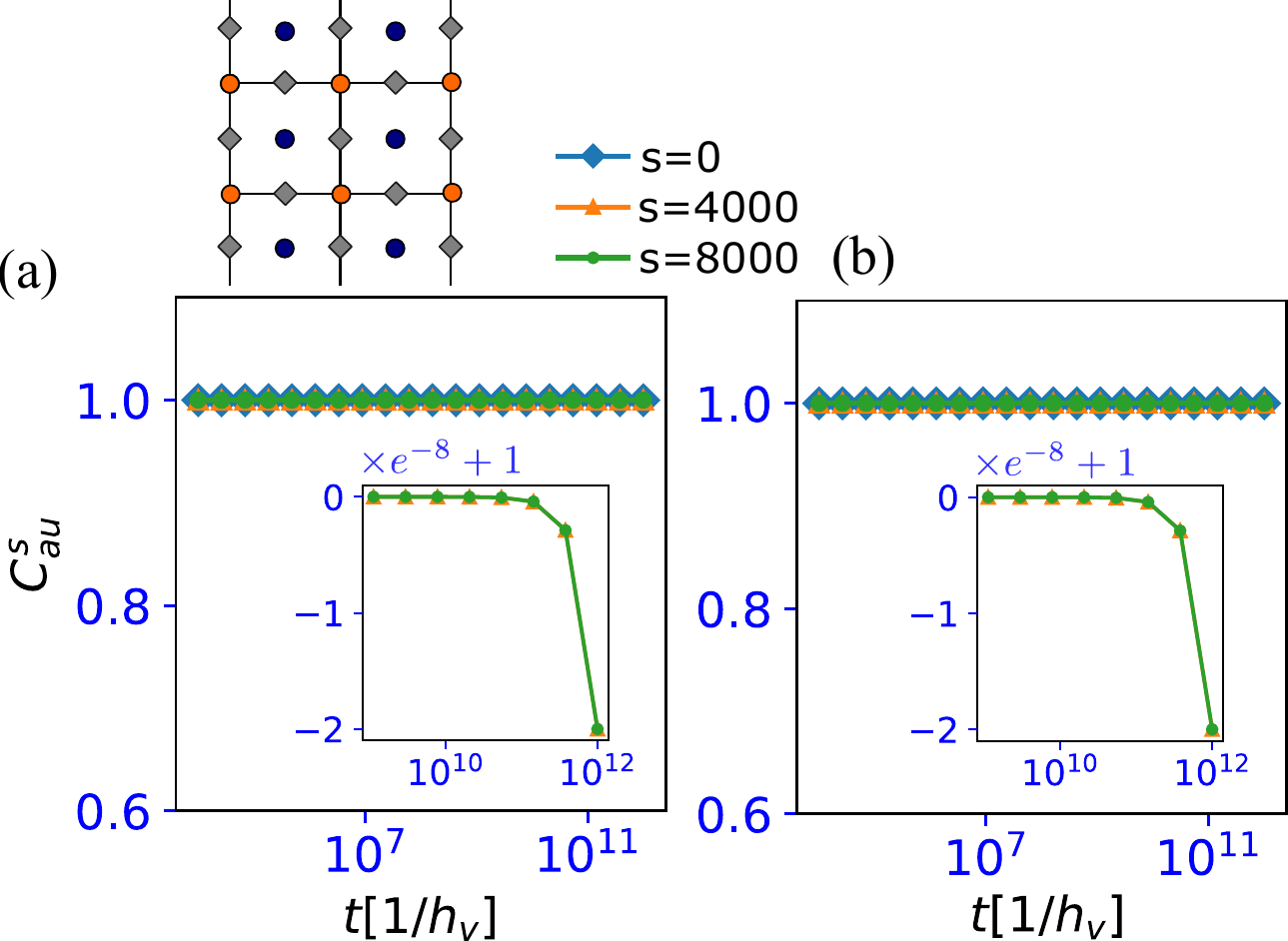}  
\end{center} 
\caption{Autocorrelator of the swap logical operator $\tilde{X}$.
(a) The case in the deep Higgs phase without $W_\gamma$ symmetry,  $h_v=-1$, $h_p=-1$, $J^z_{\ell}=-4$, $J^x_{\ell}=-0.5$, $v_1=0.5$, $v_2=0$. 
The upper figure displays the system of numerical simulation, where the total number of link is $13$. 
(b) The case in the vicinity of confinement and Higgs phase transition, $h_v=-1$, $h_p=-1$, $J^z_{\ell}=-2$, $J^x_{\ell}=-1.9$, $v_1=0.5$, $v_2=0$. 
The right inset panels for (a) and (b) show the detailed behavior for late-time evolution of the autocorrelator.
}
\label{Fig8}
\end{figure}

We study the single-encoded qubit system as shown in the inset in Fig.~\ref{Fig8} and first focus on the deep Higgs phase, 
where the logical operator $\tilde{X}$ is defined by Eq.~(\ref{logical_OP_1Q}).
In order to extract $\tilde{Z}=+1$ encoded state, we add a very small potential $v_z\tilde{Z}$ with $v_z=10^{-6}$. 
The numerically obtained time evolution is shown in Fig.~\ref{Fig8}(a), where we pick up three eigenstates, i.e.,
the ground state with $(P,S_Z,\tilde{Z})=(-1,+1,+1)$ labeled as $s=0$, a middle excited state $\tilde{Z}=+1$, $s=4000$ and a highly-excited state $\tilde{Z}=+1$, $s=8000$. 

We observe that $C^s_{\rm au}(t)$ keeps perfect coherence for a long period in all eigenstates. 
This result indicates that the degeneracy originating from the nature of  encoded qubit 
is observed in the whole energy spectrum, implying the existence of the strong zero mode. 
In addition, in the very late-time behavior of $C^s_{\rm au}(t)$, a tiny decay is observed as shown in 
the inset panel in Fig.~\ref{Fig8}(a). 
This behavior is similar to the decay of the autocorrelator in the TFIM investigated in \cite{Kemp2017}. 
However, the decay is very small, thus, we expect that the coherence of the autocorrelater in our model is almost perfect for very long times. 
This result strongly supports the expectation that the degenerate structure of encoded qubit is maintained in the whole energy spectrum of the model.

We further show the calculations of the autocorrelater in the vicinity of the confinement-Higgs phase transition in Fig.~\ref{Fig8}(b), where the parameters are set as 
$h_v=-1$, $h_p=-1$, $J^z_{\ell}=-2$, $J^x_{\ell}=-1.9$, $v_1=0.5$, $v_2=0$.
We consider eigenstates with $s=0, 4000$ and $8000$ as in the previous case in Fig~\ref{Fig8}(a). 
Even though the system is in the critical regime, the autocorrelater exhibits the same behavior with
that in deep Higgs regime.
Therefore, the numerics indicates that the encoded qubit degeneracy in the whole energy spectrum persists for the entire gauge-theoretical phase diagram.\\  

\section{DISCUSSION AND CONCLUSIONS}
In this work, we propose the subsystem-code formalism of generic LGTs. In order to exemplify the proposal, 
we studied the interplay between (2+1)-D lattice $Z_2$ extended gauge-Higgs model and subsystem code beyond the recent works \cite{Verresen2022,Wildeboer2022}.

We gave the explicit analytical description of the encoded qubit state of this subsystem code. 
The state description gives dramatic and useful understanding of gauge-theory phases, Higgs and confinement phases. 
The analytically-obtained states exhibit the boundary SSB, which is robust for the explicit breaking of the $W_\gamma$ or $H_\gamma$-symmetry in the Hamiltonian 
since these symmetries are one-form symmetry. Our work clearly shows the existence of genuine order parameters for the SSB of the global charge symmetry as well as charge confinement.

We remark that the previous study \cite{Wildeboer2022} does not give explicit form of the states 
of the subsystem code, nor discuss the connection between gauge theory and SPT state. 
The accomplishment of this work is to elucidate the strong relationship between the degeneracy and structure of the ground states in the gauge-Higgs model 
with the origin of encoded qubits in subsystem code. 
We gave some concrete analytical descriptions for not only single but also multiply-encoded qubit in both Higgs and confinement regimes.
We also constructed highly-excited eigenstates of the encoded qubits in both Higgs and confinement regimes. 
We numerically verified the above observations and investigated the spectra and state structure of the model rather in detail. 
Even though system size is small, we obtained satisfactory results for the degeneracy of encoded qubits in both Higgs and confinement regimes. 
All the numerical results are in good agreement with the analytical observations and also theoretical consideration on the (2+1)-D lattice $Z_2$ extended gauge-Higgs model. 

Furthermore, by observing the autocorrelation numerically, we found that it indicates the existence of the strong zero mode in the Higgs regime as analytical study predicts. 
The numerical results strongly support the existence of the degenerate structure of the subsystem code up to high-energy regime 
as predicted by \cite{Wildeboer2022}, and indicate that
the subsystem code is robust even at finite temperatures.

Finally, the present formalism of lattice gauge-Higgs model clarifies that the confinement phase
is an SPT phase as the Higgs phase.
SPT string order parameter is given by 
$\langle X_p \sigma^x \sigma^x \cdots \sigma^x X_{p'}\rangle \neq 0$.
 
As future interesting topic, we point out the followings: 
\begin{enumerate}
\item 
Most of the studies on gauge-theoretical systems for constructing qubits including the troic code and subsystem stabilizer code focus on the topological phase, 
which is nothing but the deconfinement phase of the LGT. The Higgs-confinement phase in the two-dimensional systems under ordinary boundary conditions is sometimes regarded as a trivial phase with only gapful excitations~\cite{Trebst2007,Dusuel2011,Else2017}. 
In the present work, however, we showed that an explicit description of the boundary states is possible and it plays a very important role in 
clarifying structure of the encoded qubits. 
From this point of view, the explicit description of the boundary states in deconfinement phase is desired to study the detailed structure of encoded qubits in 
the present model, although it might be rather complicated.

\item
We are interested in effects of disorder. 
If a kind of disorders is introduced on rough or smooth boundary, the SSB in Higgs and confinement regimes might survive up to highly-excited states even in the gauge-Higgs model.
In fact, this conjecture has been investigated from the viewpoint of many-body localization \cite{Bauer2013,Huse2013,Wahl2020}. 
Concrete numerical verification for that in certain one-dimensional models has been reported in recent works \cite{Kjall2014,Bahri2015,Decker2020,Wahl2022}. 
The confirmation of the conjecture for the LGT studied here is a future problem.

\item
We also expect that the encoded qubit state discussed in this work can be produced by a measurement-only circuit \cite{Ippoliti2021,Lavasani2021,Klocke2022,Lavasani2021_2}, where sequential projective random measurements of each terms in $H_{\rm GHM}$ are applied \cite{KI2023}. 
The numerical verification of it can be an important future direction of study 

\item Application of the present formalism to higher-dimensional models and other various types of stabilizer code Hamiltonians related to quantum memory \cite{Weinstein2019} is an interesting subject. 
Our observation seems to indicate that a finite-temperature phase transition of SSB of charge symmetry as well as confinement belong to the universality class of 
the corresponding quantum spin model in one-lower dimensions. Another interesting issue to be clarified is  meaning of the strong zero modes
from the viewpoint of the gauge-Higgs model, which are expected to exist even at (very) finite temperature. Monte-Carlo simulation is useful for these studies~\cite{Takashima2005,Ono2009}.

\item 
Finally, we comment on the experimental realization and related modification of the model of $H_{\rm GHM}$ [Eq.~(\ref{HGHM})] 
focusing on the gauge-invariant conditions. 
We are considering quantum simulation of the present models by using, e.g., ultra-cold atomic gasses, ion straps, etc.
It is a hard task for real experiment to strictly implement the gauge-invariant condition of Eq.~(\ref{const}). 
However, a promising method was proposed recently: it employs concept of local pseudogenerators (LPGs) and adds an energy penalty term of the LPGs
to the model $H_{\rm GHM}$ \cite{Halimeh2022,Homeier2023} instead of directly enforcing the Gauss's law  [Eq.~(\ref{const})]. 
The energy penalty term of the LPGs can be suitable for experimental realization of the Gauss' law by its simple form, 
and it effectively generates the target sector of the constraint of Eq.~(\ref{const}). 
We expect that in such experimentally-suitable Hamiltonian modified from $H_{\rm GHM}$, it is possible to construct the subsystem codes presented 
in this work although the toric code model $H_{\rm TC}$ [Eq.~(\ref{HTC})] as a gauge-fixing model does not appear 
and the exact solutions of the subsystem code in the modified system would be more complicated than those derived in this work. 
However, physical properties of the resultant states are essentially the same with the original ones.
Detailed investigation of this topic is interesting and a constructive future direction of study.

\end{enumerate}
\section*{Acknowledgements}
This work is supported by JSPS KAKEN-HI Grant Number 23K13026 (Y.K.). 



\begin{thebibliography}{99}
\bibitem{noteGF}
By singly applying $U_p$ to $H_{\rm GHM}$, $X_p\to 1, Z_p\to \tilde{B}_p$, and then, the standard gauge-Higgs Hamiltonian is obtained.
\bibitem{note2}
In other words, the boundary SSB of the global charge symmetry is observed through the correlation function of the gauge-invariant `spin' operator $\sigma^z_{\ell_v}Z_v$ residing on dangling links on the rough boundaries
as in the ordinary ferromagnetic SSB case, such as $\mbox{lim}_{|v-v'|\to \infty}\langle\sigma^z_{\ell_v}Z_v\sigma^z_{\ell_{v'}}Z_{v'}\rangle\neq 0$, which we observe numerically later on.
Finite value of the above correlation function means the condensation of $\sigma^z_{\ell_v}Z_v$.
\end{thebibliography}
\end{document}